\documentclass{article}
\usepackage[numbers]{natbib}
\usepackage{subcaption}
\usepackage{microtype}
\usepackage{graphicx}
\usepackage{booktabs} 
\usepackage{hyperref}

\usepackage{fullpage}
\usepackage{wrapfig}
\usepackage{hyperref,graphicx,amsmath,amssymb,bm,breakurl,epsfig,epsf,color,mathbbol,fmtcount,semtrans,multirow,comment,boldline,movie15,pgfplots}
\usepackage{tcolorbox}
\tcbuselibrary{skins}
\usepackage{tikz}
\usetikzlibrary{pgfplots.groupplots}
\usepackage[utf8]{inputenc} 
\usepackage[T1]{fontenc}    
\usepackage{url}            
\usepackage{booktabs}       
\usepackage{amsfonts}       
\usepackage{nicefrac}       
\usepackage{microtype}      
\usepackage{mathtools}
\definecolor{darkred}{RGB}{150,0,0}
\definecolor{darkgreen}{RGB}{0,150,0}
\definecolor{darkblue}{RGB}{0,0,200}
\hypersetup{colorlinks=true, linkcolor=darkred, citecolor=darkgreen, urlcolor=darkblue}

\newtheorem{assumption}{Assumption}

\numberwithin{equation}{section}

\newcommand{\cln}[1]{\textcolor{red}{}}

\def \endprf{\hfill {\vrule height6pt width6pt depth0pt}\medskip}

\newcommand{\tsn}[1]{{\left\vert\kern-0.25ex\left\vert\kern-0.25ex\left\vert #1 
		\right\vert\kern-0.25ex\right\vert\kern-0.25ex\right\vert}}

\newcommand{\beq}{\begin{equation}}
	\newcommand{\ba}{\begin{align}}
		\newcommand{\ea}{\end{align}}

	\newcommand{\eeq}{\end{equation}}

\newcommand{\opnorm}[1]{\left\|#1\right\|}

\newcommand{\abs}[1]{\left|#1\right|}

\definecolor{emmanuel}{RGB}{255,127,0}

\newcommand{\vct}[1]{\bm{#1}}
\newcommand{\mtx}[1]{\bm{#1}}

\newcommand{\FT}{{\mathcal{F}}}
\newcommand{\IFT}{{\mathcal{F}^{-1}}}
\newcommand{\mask}{{\mtx{M}}}
\newcommand{\fwd}[1]{{\mathcal{A}\left(#1\right) }}
\DeclareMathOperator*{\argmin}{arg\,min}

\begin{document}
\onecolumn
\title{Data augmentation for deep learning based accelerated MRI reconstruction with limited data}
\author{Zalan Fabian\thanks{{Dept. of Electrical and Computer Eng., University of Southern California, Los Angeles, CA.~~~~~Email: \url{zfabian@usc.edu}}}\quad\quad\quad Reinhard Heckel\thanks{{Dept. of Electrical and Computer Eng., Technical University of Munich, Munich, Germany, and Dept. of Electrical and Computer Eng., Rice University.}~~~~~~~~~Email: \url{reinhard.heckel@tum.de}}\quad\quad\quad Mahdi Soltanolkotabi\thanks{Dept. of Electrical and Computer Eng., University of Southern California, Los Angeles, CA.~~~~~~Email: \url{soltanol@usc.edu}}}
\maketitle

{\let\thefootnote\relax\footnotetext{This work will appear in the International Conference on Machine Learning (ICML) 2021.}}
\begin{abstract}
	Deep neural networks have emerged as very successful tools for image restoration and reconstruction tasks. 
	These networks are often trained end-to-end to directly reconstruct an image from a noisy or corrupted measurement of that image. 
	To achieve state-of-the-art performance, training on large and diverse sets of images is considered critical.
	However, it is often difficult and/or expensive to collect large amounts of training images. Inspired by the success of Data Augmentation (DA) for classification problems, in this paper, we propose a pipeline for data augmentation for 
	accelerated MRI reconstruction
	and study its effectiveness at reducing the required training data in a variety of settings. 
	Our DA pipeline, MRAugment, is specifically designed to utilize the invariances present in medical imaging measurements as naive DA strategies that neglect the physics of the problem fail. 
	Through extensive studies on multiple datasets we demonstrate
	that in the low-data regime DA prevents overfitting and can match or even surpass the state of the art while using significantly fewer training data, whereas in the high-data regime it has diminishing returns. 
	Furthermore, our findings show that DA can improve the robustness of the model against various shifts in the test distribution.
\end{abstract}

\section{Introduction}

In magnetic resonance imaging (MRI), an extremely popular medical imaging technique, it is common to reduce the acquisition time by subsampling the measurements, because this reduces cost and increases accessibility of MRI to patients. 
Due to the subsampling, there are fewer equations than unknowns, and therefore the signal is not uniquely identifiable from the measurements. To overcome this challenge there has been a flurry of activity over the last decade aimed at utilizing prior knowledge about the signal, in a research area referred to as \textit{compressed sensing} \citep{cs:candesStableSignalRecovery2006, cs:donohoCompressedSensing2006}. 

Compressed sensing methods reduce the required number of measurements by utilizing prior knowledge about the images during the reconstruction process, traditionally via a convex regularization that enforces sparsity in an appropriate transformation of the image. More recently, deep learning techniques have been used to enforce much more nuanced forms of prior knowledge (see \citet{inv:ongieDeepLearningTechniques2020} and references therein for an overview). The most successful of these approaches aim to directly learn the inverse mapping from the measurements to the image by training on a large set of training data consisting of signal/measurement pairs. This approach often enables  faster reconstruction of images, but more importantly, deep learning techniques yield significantly higher quality reconstructions. Thus, deep learning techniques enable reconstructing a high-quality image from fewer measurements which further reduces image acquisition times. For instance, in an accelerated MRI competition known as fastMRI Challenge \citep{ fastmri:zbontar2018fastmri}, 
all the top contenders used deep learning reconstruction techniques.

Contrary to classical compressive sensing approaches, however, deep learning techniques typically rely on large sets of training data consisting of images along with the corresponding measurement. This is also true about the use of deep learning techniques in other areas such as computer vision and Natural Language Processing (NLP) where superb empirical success has been observed. 
While large datasets have been harvested and carefully curated in areas such as vision and NLP, this is not feasible in many scientific applications including MRI. It is difficult and expensive to collect the necessary datasets for a  variety of reasons, including patient confidentiality requirements, cost and time of data acquisition, lack of medical data compatibility standards, and the rarity of certain diseases. 

A common strategy to reduce reliance on training data in classification tasks is data augmentation. Data augmentation techniques are used in classification tasks to significantly increase the performance on standard benchmarks such as ImageNet and CIFAR-10. For a comprehensive survey of image data augmentation in deep learning see \cite{da:shorten2019survey}. More specific to medical imaging, data augmentation techniques have been successfully applied to registration, classification and segmentation of medical images. Recently, several studies \citep{ganda:zhaoImageAugmentationsGAN2020, ganda:karras2020training, ganda:zhaoDifferentiableAugmentationDataEfficient2020} have demonstrated that data augmentation can significantly reduce the data needed for GAN training for high quality image generation. 
In a classification setting, data augmentation consists of adding additional synthetic data obtained by performing \emph{invariant} alterations to the data (e.g.~flips, translations, or rotations) which do not affect the response (i.e., the label). 

In image reconstruction tasks, however, data augmentation techniques are less common and much more difficult to design because the response (the measurement) is affected by the data augmentation. For example, measurements of a rotated image are not the same as measurements from the original image. In the context of accelerated MRI reconstruction, augmentation techniques such as randomly generated undersampling masks \citep{da:liu2019santis} and simple random flipping \citep{da:lee2018deep} have been applied, and authors in \citet{da:schlemper2017deep} note the importance of rigid transforms in avoiding overfitting. However, an effective pipeline of augmentations for training data reduction and thorough experimental studies thereof are lacking. 


The goal of this paper is to explore the benefits of data augmentation techniques for accelerated MRI with limited training data. By carefully taking into account the physics behind the MRI acquisition process we design a data augmentation pipeline, which we call MRAugment \footnote{Code: \url{https://github.com/MathFLDS/MRAugment}}, that can successfully reduce the amount of training data required. Our contributions are as follows:
\begin{itemize}
	\item We perform an extensive empirical study of data augmentation in accelerated MRI reconstruction. To the best of our knowledge, this work is the first in-depth experimental investigation focusing on the benefits of data augmentation in the context of training data reduction for accelerated MRI.
	\item We propose a data augmentation technique tailored to the physics of the MR reconstruction problem. It is not obvious how to perform data augmentation in the context of accelerated MRI or in inverse problems in general, because by changing an image to enlarge the training set, we do not automatically get a corresponding measurement, contrary to classification problems, where the label is retained.
	\item We demonstrate the effectiveness of MRAugment on a variety of datasets. On small datasets we report significant improvements in reconstruction performance on the full dataset when MRAugment is applied. Moreover, on small datasets we are able to surpass full dataset baselines by using only a small fraction of the available training data by leveraging our proposed data augmentation technique. 
	\item We perform an extensive study of MRAugment on a large benchmark accelerated MRI data set, specifically on the fastMRI \citep{fastmri:zbontar2018fastmri} dataset. 
	For 8-fold acceleration and multi-coil measurements 
	(multi-coil measurements are the standard acquisition mode for clinical practice)  
	we achieve performance on par with the state of the art with only $10\%$ of the training data.
	Similarly, again for 8-fold acceleration and single-coil experiments (an acquisition mode popular for experimentation) MRAugment can achieve the performance of reconstruction methods trained on the entire dataset while using only 33\% of training data. 
	\item We reveal additional benefits of data augmentation on model robustness in a variety of settings. We observe that MRAugment has the potential to improve generalization to unseen MRI scanners, field strengths and anatomies. Furthermore, due to the regularizing effect of data augmentation, MRAugment prevents overfitting to training data and therefore may help eliminate hallucinated features on reconstructions, an unwanted side-effect of deep learning based reconstruction.
	%
	
\end{itemize}

\section{Background and Problem Formulation}


MRI is a medical imaging technique that exploits strong magnetic fields to form images of the anatomy. MRI is a prominent imaging modality in diagnostic medicine and biomedical research because it does not expose patients to ionizing radiation, contrary to competing technologies such as computed and positron emission tomography. 

However, performing an MR scan is time intensive, which is problematic for the following reasons. 
First, patients are exposed to long acquisition times in a confined space with high noise levels. 
Second, long acquisition times induce reconstruction artifacts caused by patient movement, which sometimes requires patient sedation in particular in pediatric MRI \citep{intro:vasanawala2010improved}. 
Reducing the acquisition time can therefore increase both the accuracy of diagnosis and patient comfort. Furthermore, decreasing the acquisition time needed allows more patients to receive a scan using the same machine. This can significantly reduce patient cost, since each MRI machine comes with a high cost to maintain and operate.

Since the invention of MR in the $1980$s there has been tremendous research focusing on reducing their acquisition time. 
The two main ideas are to i) perform multiple acquisitions simultaneously \citep{parallel:sodickson1997simultaneous, parallel:pruessmann1999sense, parallel:griswold2002generalized} and to ii) subsample the measurements, known as accelerated acquisition or compressed sensing \citep{csmri:lustigCompressedSensingMRI2008b}. Most modern scanners combine both techniques, and therefore we consider such a setup.


\subsection{Accelerated MRI acquisition}
In magnetic resonance imaging, measurements of a patient's anatomy are acquired in the Fourier-domain, also called \textit{k-space}, through receiver coils. In the \textit{single-coil} acquisition mode, the k-space measurement $\vct{k} \in \mathbb{C}^n$ of a complex-valued ground truth image $\vct{x}^*\in \mathbb{C}^n$ is given by
\begin{equation*}
	\vct{k} = \FT \vct{x}^* + \vct{z},
\end{equation*}
where $\FT$ is the two-dimensional Fourier-transform, and $\vct{z} \in \mathbb{C}^n$ denotes additive noise arising in the measurement process. 
In parallel MR imaging, multiple receiver coils are used, each of which captures a different region of the image, represented by a complex-valued sensitivity map $\mtx{S}_i$. In this \textit{multi-coil} setup, coils acquire k-space measurements modulated by their corresponding sensitivity maps:
\begin{equation*}
	\vct{k}_i = \FT \mtx{S}_i \vct{x}^* + \vct{z}_i, \quad i=1, .., N,
\end{equation*}
where $N$ is the number of coils. 
Obtaining fully-sampled k-space data is time-consuming, and therefore in accelerated MRI we decrease the number of measurements by undersampling in the Fourier-domain. This undersampling can be represented by a binary mask $\mask$ that sets all frequency components not sampled to zero:
\begin{equation*}
	\vct{\tilde{k}}_i = \mask \vct{k}_i ,~~ i=1, .., N.
\end{equation*}
We can write the overall forward map concisely as 
\[
\vct{\tilde{k}} = \fwd{\vct{x}^*},
\]
where $\fwd{\cdot}$ is the linear forward operator and $\vct{\tilde{k}}$ denotes the undersampled coil measurements stacked into a single column vector. 
The goal in accelerated MRI reconstruction is to recover the image $\vct{x}^*$ from the set of k-space measurements $\vct{\tilde{k}}$. 
Note that---without making assumptions on the  image $\vct{x}^*$---it is in general impossible to perfectly recover the image, because we have fewer measurements than variables to recover. This recovery problem is known as compressed sensing. 
To make image recovery potentially possible, 
recovery methods make structural assumptions about $\vct{x}^*$, such that it is sparse in some basis or implicitly that it looks similar to images from the training set. 
%
\subsection{Traditional accelerated MRI reconstruction methods}
Traditional compressed sensing recovery methods for accelerated MRI are based on assuming that the image $\vct{x}^*$ is sparse in some dictionary, for example the wavelet transform. Recovery is then posed typically as a convex optimization problem:
\begin{equation*}
	\vct{\hat{x}} 
	= \argmin_{\vct{x}} \opnorm{  \fwd{\vct{x}} - \vct{\tilde{k}} }^2 + \mathcal{R}(\vct{x}),
\end{equation*}
where $\mathcal{R}(\cdot)$ is a regularizer enforcing sparsity in a certain domain. Typical functions used in CS based MRI reconstruction are $\ell_1$-wavelet and total-variation regularizers. These optimization problems can be numerically solved via iterative gradient descent based methods.

\subsection{Deep learning based MRI reconstruction methods}
In recent years, several deep learning algorithms have been proposed and convolutional neural networks established new state of the art in MRI reconstruction significantly surpassing the classical baselines. Encoder-decoder networks such as the U-Net \citep{unet:ronneberger2015u} and its variants were successfully used in various medical image reconstruction \citep{unet:hyun2018deep, unet:han2018framing} and segmentation problems \citep{unet:cciccek20163d, unet:zhou2018unet++}. These models consist of two sub-networks:  the encoder repeatedly filters and downsamples the input image with learned convolutional filters resulting in a concise feature vector. This low-dimensional representation is then fed to the decoder consisting of subsequent upsampling and learned filtering operations.  Another approach that can be considered a generalization of iterative compressed sensing reconstructions consists of unrolling the data flow graph of popular algorithms such as ADMM \citep{unroll:yang2016deep} or gradient descent iterations \citep{istanet:zhang2018ista} and mapping them to a cascade of sub-networks.  Several variations of this unrolled method have been proposed recently for MR reconstruction, such as i-RIM \citep{irim:putzky2019invert}, Adaptive-CS-Net \citep{csnet:pezzotti2019adaptive}, Pyramid Convolutional RNN \citep{pyrconv:wang2019pyramid} and E2E VarNet \citep{varnet:sriram2020end}.

Another line of work, inspired by the deep image prior~\citep{ulyanov_DeepImagePrior_2018} focuses on using the inductive bias of convolutional networks to perform reconstruction without any training data~\citep{jin_TimeDependentDeepImage_2019,darestani_CanUntrainedNeural_2020,heckel_CompressiveSensingUntrained_2020,heckel_DeepDecoderConcise_2019,untrained:van2018compressed}. Those methods do perform significantly better than classical un-trained networks, but do not perform as well as neural networks trained on large sets of training data.

\section{MRAugment: a data augmentation pipeline for MRI}

In this section we propose our data augmentation technique, MRAugment, for MRI reconstruction. 
We emphasize that data augmentation in this setup and for inverse problems in general is substantially different from DA for classification problems.
For classification tasks, the label of the augmented image is trivially the same as that of the original image, whereas for inverse problems we have to generate both an augmented target image and the corresponding measurements. 
This is non-trivial as it is critical to match the noise statistics of the augmented measurements with those in the dataset.

We are given training data in the form of fully-sampled MRI measurements in the Fourier domain, and our goal is to generate new training examples consisting of a subsampled k-space measurement along with a target image.
MRAugment is model-agnostic in that the generated augmented training example can be used with any machine learning model and therefore can be seamlessly integrated with existing reconstruction algorithms for accelerated MRI, and potentially beyond MRI.

\begin{figure}[t]
	\centering
	\includegraphics[width=0.8\linewidth]{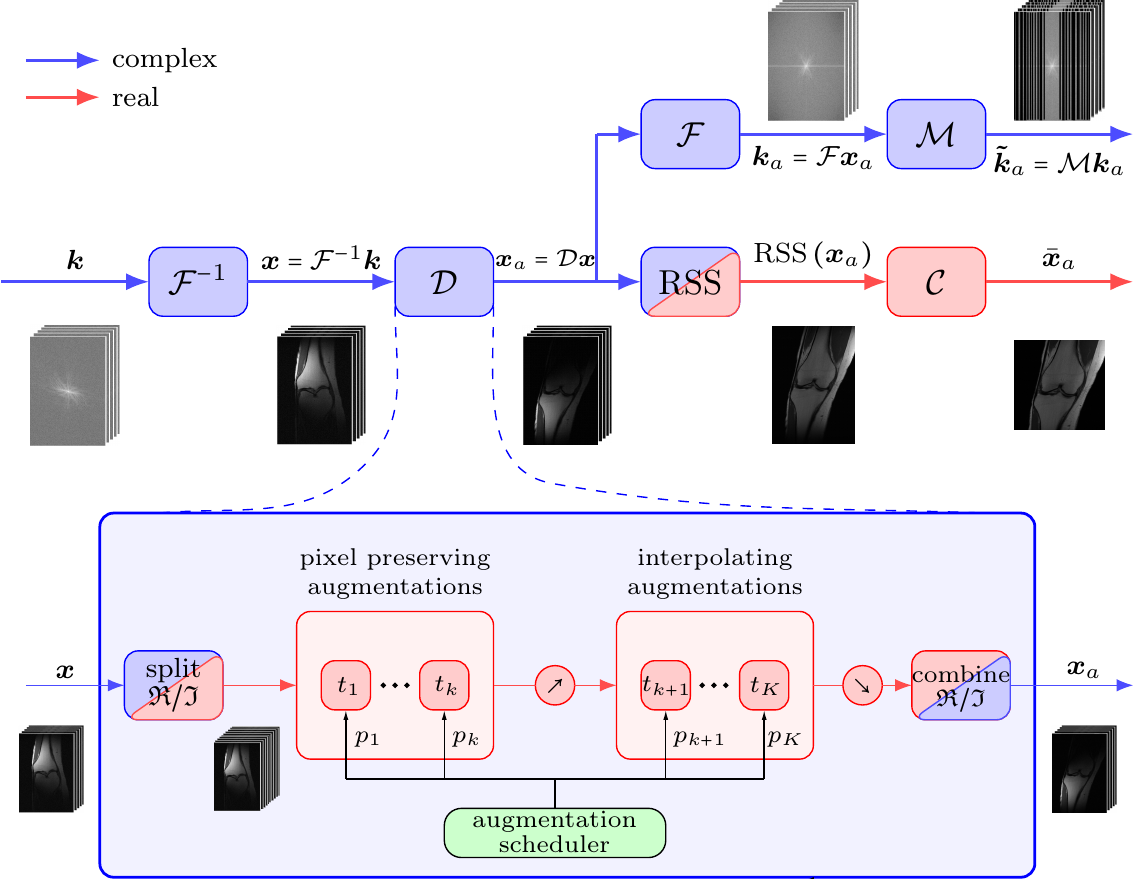}
	\caption{	\label{fig:da_flowchart}Flowchart of MRAugment, our data augmentation pipeline for MRI.}
\end{figure}

Our data augmentation pipeline, illustrated in Figure \ref{fig:da_flowchart}, generates a new example consisting of a subsampled k-space measurement $\vct{\tilde k}_a$ along with a target image $\vct{\bar x}_a$ as follows. We are given training examples as fully-sampled k-space slices, which we stack into a single column vector $\vct{k} = \text{col}(\vct{k}_1, \vct{k}_2, ..., \vct{k}_N)$ for notational convenience. From these, we obtain the individual coil images by applying the inverse Fourier transform as $\vct{x} = \mathcal{F}^{-1}\vct{k}$. 
We generate augmented individual coil images with an augmentation function $\mathcal{D}$, specified later, as $\vct{x}_a = \mathcal{D}(\vct{x})$.
From the augmented images, we generate an undersampled measurement by applying the forward model as $\vct{\tilde{k}}_a = \fwd{ \vct{x}_a }$.  
Both $\vct{x}$ and $\vct{x}_a$ are \textit{complex-valued}: even though the MR scanner obtains measurements of a real-valued object, due to noise the inverse Fourier-transform of the measurement is complex-valued.
Therefore the augmentation function has to generate complex-valued images, which adds an extra layer of difficulty compared to traditional data augmentation techniques pertaining to real-valued images 
(see Section \ref{sec:noise} for further details). 
Finally, the real-valued ground truth image is 
obtained by combining the coil images $\vct{x}_{a,i}$ by pixel-wise root-sum-squares (RSS) followed by center-cropping $\mathcal C$:
\begin{align*}
	{\vct{\bar x}_a} =\mathcal{C} \left(\text{RSS}(\vct{x}_a)\right) = \mathcal{C} \left(\sqrt{\sum_{i=1}^N \abs{\vct{x}_{a,i}}^2}\right).
\end{align*}
In the following subsections we first argue why we generate individual coil images with the augmentation function, then discuss the design of the augmentation function $\mathcal D$ itself.

\subsection{
	Data augmentation needs to preserve noise statistics
	\label{sec:noise}
}
As mentioned before, we are augmenting complex-valued, noisy images. This noise enters in the measurement process when we obtain the fully-sampled measurement of an image $\vct{x}^\ast$ as $\vct{k} = \FT\vct{x}^* + \vct{z}$, and is well approximated by i.i.d complex Gaussian noise, independent in the real and imaginary parts of each pixel \citep{mribook:nishimura1996principles}. Therefore, we can write $\vct{x} = \vct{x}^* + \vct{z}'$ where $\vct{z}'$ has the same distribution as $\vct{z}$ due to $\FT$ being unitary. Since the noise distribution is characteristic to the instrumental setup (in this case the MR scanner and the acquisition protocol), assuming that the training and test images are produced by the same setup, it is important that the augmentation function preserves the noise distribution of training images as much as possible. Indeed, a large mismatch between training and test noise distribution leads to poor generalization \citep{snr:knollAssessmentGeneralizationLearned2019}.

Let us demonstrate why it is non-trivial to generate augmented measurements for MRI through a simple example. A natural but perhaps naive approach for data augmentation is to augment the real-valued target image $\bar{\vct{x}}$ instead of the complex valued $\vct{x}$. 
This would allow us to directly obtain real augmented images from a real target image just as in typical data augmentation. However, this approach leads to different noise distribution in the measurements compared to the test data due to the non-linear mapping from individual coil images to the real-valued target and works poorly. Experiments demonstrating the weakness of this naive approach of data augmentation can be found in Section \ref{sec:exp_noise}.


In contrast, if we augment the individual coil images $\vct{x}$ directly with a linear function $\mathcal{D}$, which is our main focus here, we obtain the augmented k-space data $$\vct{k}_a =  \FT \mathcal{D}\vct{x} = \FT\mathcal{D} (\vct{x}^* + \vct{z}') = \FT\mathcal{D}\vct{x}^* +  \FT\mathcal{D}\vct{z}',$$ where $\FT\mathcal{D}\vct{x}^*$ represents the augmented signal and the noise $\FT\mathcal{D}\vct{z}'$ is still additive complex Gaussian. A key observation is that in case of transformations such as translation, horizontal and vertical flipping and rotation the noise distribution is exactly preserved. Moreover, for general linear transformations the noise is still Gaussian in the real and imaginary parts of each pixel. 

To elaborate further, in the multi-coil case our augmentation pipeline applies transformations to the underlying object modulated by the different coil sensitivity maps. In particular, the fully sampled measurement of the $i$th coil in the image domain takes the form 
\begin{equation}\label{eq:sens}
	x_i = S_i x^* + z_i',
\end{equation}
where  $z_i' = \IFT z_i$ is i.i.d Gaussian noise obtained via a unitary transform of the original measurement noise. Assuming linear augmentations, the augmented coil image from MRAugment can be written as
\begin{equation} \label{eq:sens_1}
	x_{a,i} = \mathcal{D}(S_i x^* + z_i')=\mathcal{D}S_i x^* + \mathcal{D}z_i',
\end{equation}  
where the additive noise is still Gaussian. As seen in \eqref{eq:sens_1}, MRAugment transforms images modulated by the coil sensitivities, therefore the sensitivitiy maps are also indirectly augmented. However, the models we experimented with had no issues learning the proper mapping from augmented measurements with transformed sensitivity maps as our experimental results show.

It is natural to ask if data augmentation would be possible by directly augmenting the object, before the coil sensitivities are applied. If the sensitivity maps are known or are estimated a priori, one may recover the object from the various coils as 
\begin{align*}
	x = \sum_{j=1}^N S_j^* x_j = \sum_{j=1}^N S_j^*(S_j x^* + z_j') =  x^* + \sum_{j=1}^N S_j^* z_j', 
\end{align*}  
where $S_j^*$ is the complex conjugate of $S_j$ and $\sum_{j=1}^N S_j^*S_j = I$ due to typical normalization \citep{varnet:sriram2020end}.
Then, we can apply the augmentation as
\begin{equation*}
	x_a = \mathcal{D}x = \mathcal{D}(x^* + \sum_{i=1}^N S_j^* z_j') =  \mathcal{D}x^* +  \mathcal{D}\sum_{j=1}^N S_j^* z_j'.
\end{equation*}  
Finally, we obtain the augmented coil images as
\begin{equation} \label{eq:sens_2}
	x_{a,i} = S_i x_a = S_i \mathcal{D}x^* + S_i \mathcal{D}\sum_{j=1}^N S_j^* z_j'.
\end{equation} 
Comparing \eqref{eq:sens_2} with \eqref{eq:sens_1}, one may see that now the augmentation is directly applied to the ground truth signal bypassing the coil sensitivities. However, comparing this result in \eqref{eq:sens_2} with the original unaugmented coil images in \eqref{eq:sens} reveals that the additive noise in \eqref{eq:sens_2} has a very different distribution from the original i.i.d Gaussian, even worse noise on different augmented coil images are now correlated. 
Finally, the sensitivity maps are typically not known and need to be estimated before we can apply this augmentation technique, which can introduce additional inaccuracies in the augmentation pipeline. 

This discussion motivates our choice to i) augment complex-valued images directly derived from the original k-space measurements, ii) consider simple transformations which preserve the noise distribution and iii) augment individual coil images as in \eqref{eq:sens_1}. Next we overview the types of augmentations we propose in line with these key observations.


\subsection{Transformations used for data augmentation}
We apply the following two types of image transformations $\mathcal{D}$ in our data augmentation pipeline:\\
\noindent \textbf{Pixel preserving augmentations}, that do not require any form of interpolation and simply result in a permutation of pixels over the image. Such transformations are vertical and horizontal flipping, translation by integer number of pixels and rotation by multiples of $90^\circ$. As we pointed out in Section \ref{sec:noise}, these transformations do not affect the noise distribution on the measurements and therefore are suitable for problems where training and test data are expected to have similar noise characteristics.\\
\noindent\textbf{General affine augmentations}, that can be represented by an affine transformation matrix and in general require resampling the transformed image at the output pixel locations. Augmentations in this group are: translation by arbitrary (not necessarily integer) coordinates, arbitrary rotations, scaling and shearing. Scaling can be applied along any of the two spatial dimensions. We differentiate between isotropic scaling, in which the same scaling factor $s$ is applied in both directions ($s>1$: \textit{zoom-in}, $s<1$: \textit{zoom-out}) and anisotropic scaling in which different scaling factors $(s_x, s_y)$ are applied along different axes. 

\begin{figure}[htb]
	\captionsetup[subfigure]{labelformat=empty}
	\centering 
	\begin{subfigure}[t]{0.19\textwidth}
		\includegraphics[width=\linewidth]{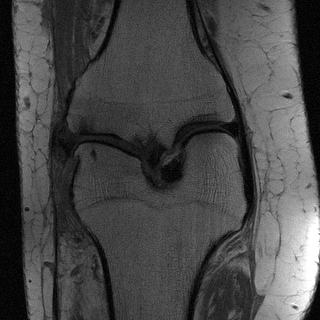}
		\caption{Original}
		\label{fig:1}
	\end{subfigure}\hfil 
	\begin{subfigure}[t]{0.19\textwidth}
		\includegraphics[width=\linewidth]{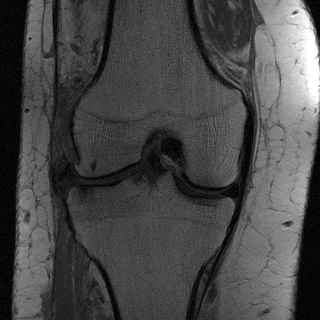}
		\caption{V-flip}
		\label{fig:2}
	\end{subfigure}\hfil 
	\begin{subfigure}[t]{0.19\textwidth}
		\includegraphics[width=\linewidth]{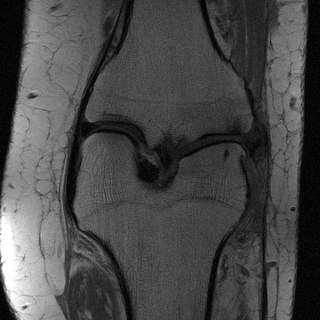}
		\caption{H-flip}
		\label{fig:3}
	\end{subfigure}\hfil 
	\begin{subfigure}[t]{0.19\textwidth}
		\includegraphics[width=\linewidth]{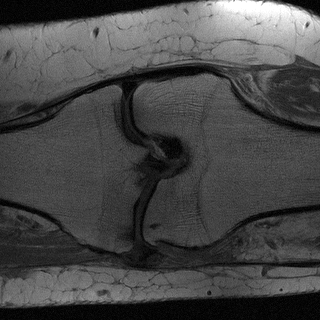}
		\caption{Rot. $k 90 ^{\circ}$}
		\label{fig:4}
	\end{subfigure}\hfil 
	\begin{subfigure}[t]{0.19\textwidth}
		\includegraphics[width=\linewidth]{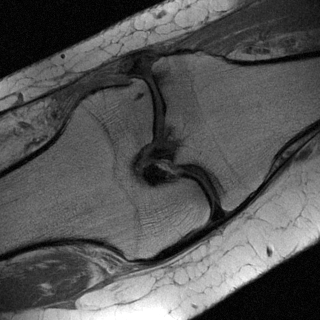}
		\caption{Rotation}
		\label{fig:5}
	\end{subfigure}\hfil 
	\begin{subfigure}[t]{0.19\textwidth}
		\includegraphics[width=\linewidth]{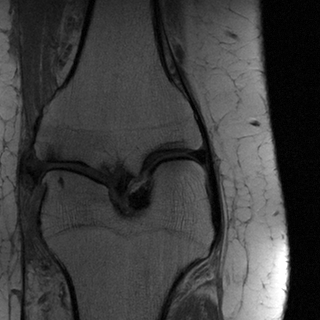}
		\caption{Transl.}
		\label{fig:10}
	\end{subfigure}\hfil 
	\begin{subfigure}[t]{0.19\textwidth}
		\includegraphics[width=\linewidth]{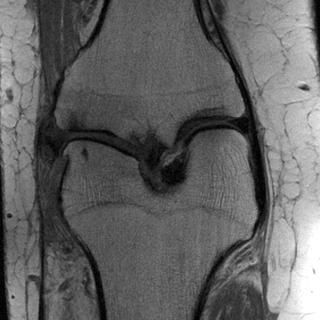}
		\caption{Zoom-in}
		\label{fig:6}
	\end{subfigure}\hfil 
	\begin{subfigure}[t]{0.19\textwidth}
		\includegraphics[width=\linewidth]{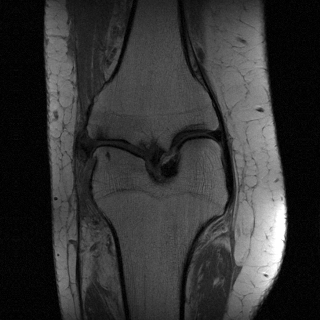}
		\caption{Zoom-out}
		\label{fig:7}
	\end{subfigure}\hfil 
	\begin{subfigure}[t]{0.19\textwidth}
		\includegraphics[width=\linewidth]{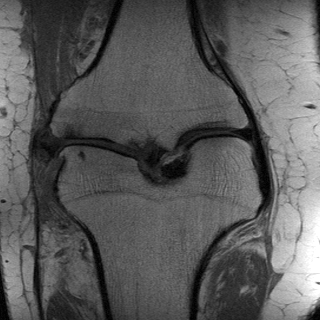}
		\caption{Aniso sc.}
		\label{fig:8}
	\end{subfigure}\hfil 
	\begin{subfigure}[t]{0.19\textwidth}
		\includegraphics[width=\linewidth]{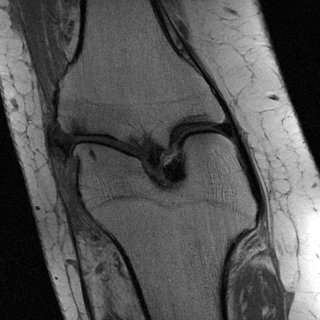}
		\caption{Shearing}
		\label{fig:9}
	\end{subfigure}
	\caption{	\label{fig:da_types}Transformations used in MRAugment applied to a ground truth slice.
		\vspace{-0.5cm}}
	
\end{figure}

Figure \ref{fig:da_types} provides a visual overview of the types of augmentations applied in this paper. Numerous other forms of transformations may be used in this framework such as exposure and contrast adjustment, image filtering (blur, sharpening) or image corruption (cutout, additive noise). However, in addition to the noise considerations mentioned before that have to be taken into account, some of these transformations are difficult to define for complex-valued images and may have subtle effects on image statistics. For instance, brightness adjustment could be applied to the magnitude image, the real part only or both real and imaginary parts, with drastically different effects on the magnitude-phase relationship of the image. 
That said, we hope to incorporate additional augmentations in our pipeline in the future after a thorough study of how they affect the noise distribution.

\subsection{Scheduling and application of data augmentations}
With the different components in place we are now ready to discuss the scheduling and application of the augmentations, as depicted in the bottom half of Figure \ref{fig:da_flowchart}. 
Recall that MRAugment generates a 
target image $\bar{\vct{x}}_a$ and corresponding undersampled k-space measurement $\vct{\tilde{k}}_a$ from a full k-space measurement. 
Which augmentation is applied and how frequently is determined by a parameter $p$, the common parameter determining the probability of applying a transformation to the ground truth image during training, and the weights $\mathcal{W} = (w_1, w_2,..., w_K)$ pertaining to the $K$ different augmentations, controlling the weights of transformations relative to each other. We apply a given transformation $t_i$ with probability $p_i = p \cdot w_i$. 
The augmentation function is applied to the coil images, specifically the same transformation is applied with the same parameters to the real and imaginary parts $(\Re\{\vct{x}_1\}, \Im\{\vct{x}_1\}, \Re\{\vct{x}_2\}, \Im\{\vct{x}_2\}, ..., \Re\{\vct{x}_N\}, \Im\{\vct{x}_N\})$ of coil images.
If a transformation $t_i$ is sampled (recall that we select them with probabilities $p_i$), we randomly select the parameters of the transformation from a pre-defined range (for example, rotation angle in $[0, 180^\circ]$). 
To avoid aliasing artifacts, we first upsample the image before transformations that require interpolation. Then the result is downsampled to the original size. 

A critical question is how to schedule $p$ over training in order to obtain the best model. 
Intuitively, in initial stages of training no augmentation is needed, since the model can learn from the available original training examples. 
As training progresses the network learns to fit the original data points and their utility decreases over time. 
We find schedules starting from $p = 0$ and increasing over epochs to work best in practice. The ideal rate of increase depends on both the model size and amount of available training data.  

\section{Experiments \label{sec:exp}}
In this section we explore the effectiveness of MRAugment in the context of accelerated MRI reconstruction in various regimes of available training data sizes on various datasets. 
We start with providing a summary of our main findings, followed by a detailed description of the experiments. Additional reconstructions and more experimental details can be found in the supplementary material.

\textbf{In the low-data regime} (up to $\approx 4k$ images), data augmentation very significantly boosts reconstruction performance. The improvement is large both in terms of raw SSIM and visual reconstruction quality. Using MRAugment, fine details are recovered that are completely missing from reconstructions without DA. This suggests that DA improves the value of reconstructions for medical diagnosis, since health experts typically look for small features of the anatomy. This regime is especially important in practice, since large public datasets are extremely rare.

\textbf{In the moderate-data regime} ( $\approx 4k - 15k$ images) MRAugment still achieves significant improvement in reconstruction SSIM. We want to emphasize the significance of seemingly small differences in SSIM close to the state of the art and invite the reader to visit the fastMRI Challenge Leaderboard that demonstrates how close the best performing models are.

\textbf{In the high-data regime} (more than $15k$ images) data augmentation has diminishing returns. It does not notably improve performance of the current state of the art, but it does not degrade performance either. Our experiments in the latter two regimes however strongly suggest that data augmentation combined with much larger models may lead to significant improvement over the state of the art, even in the high-data regime.
However, without larger models it is expected that in a regime of abundant data, DA does not improve performance. For the models and problem considered here, this is around $15k$ images. We hope to investigate the effectiveness of MRAugment combined with such larger models in our future work.

\textbf{Additional benefits} of data augmentation include improved robustness under shifts in test distribution, such as improved  generalization to new MRI scanners and field strengths. Furthermore, we observe that data augmentation can help to eliminate hallucinations by preventing overfitting to training data.

\subsection{Experimental setup} 
We use the state-of-the-art End-to-End VarNet model \citep{varnet:sriram2020end}, which is as of now one of the best performing neural network models for MRI reconstruction. 
We measure performance in terms of the structural similarity index measure (SSIM), which is a standard evaluation metric for medical image reconstruction. We study the performance of MRAugment as a function of the size of the training set. We construct different subsampled training sets by randomly sampling volumes of the original training dataset and adding all slices of the sampled volumes to the new subsampled dataset. For all experiments, we apply random masks by undersampling whole k-space lines in the phase encoding direction by a factor of $8$ and including $4\%$ of lowest frequency adjacent k-space lines in order to be consistent with baselines in \citep{fastmri:zbontar2018fastmri}. For both the baseline experiments and for MRAugment, we generate a new random mask for each slice on-the-fly while training by uniformly sampling k-space lines, but use the same fixed mask for each slice within the same volume on the validation set (different across volumes). This technique is standard for models trained on the fastMRI dataset and not specific to our data augmentation pipeline. For augmentation probability scheduling we use 
\begin{equation*}
	p(t) = \frac{p_{max}}{1-e^{-c}}(1-e^{-t c/T}),
\end{equation*}
where $t$ is the current epoch,  $T$ denotes the total number of epochs, $c = 5$ and $p_{max} = 0.55$ unless specified otherwise. This schedule works resonably well on datasets of various size that we have studied and has not been fine-tuned to individual experiments. Ablation studies on the effect of the scheduling function is deferred to the supplementary.

\subsection{Low-data regime}

For the low-data regime we work with two different datasets, the Stanford 2D FSE dataset and the 3D FSE Knee dataset described below and demonstrate significant gains in reconstruction performance. 

\textbf{ Stanford 2D FSE dataset.} First, we perform experiments on the Stanford 2D FSE \citep{website:Stanford2D} dataset, a public dataset of $89$ fully-sampled MRI volumes of various anatomies including lower extremity, pelvis and more. We use $80\% - 20\%$ training-validation split, randomly sampled by volumes. We generate $5$ random splits in order to minimize variations in reconstruction metrics due to validation set selection and report the mean validation SSIM over $5$ runs along with the standard errors.

\begin{figure}[h!]
	\centering
	\begin{subfigure}[t]{0.45\textwidth}
		\includegraphics[width=\linewidth]{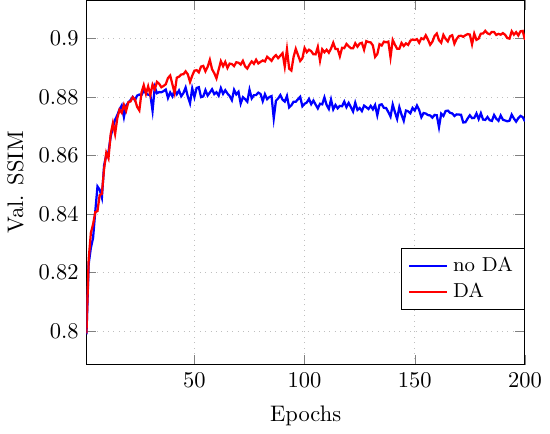}
		\caption{Validation SSIM vs. epochs. MRAugment prevents the model from overfitting to training data.}
		\label{fig:stanford2d_ssim_epochs}
	\end{subfigure}\hfil 
	\begin{subfigure}[t]{0.43\textwidth}
		\includegraphics[width=\linewidth]{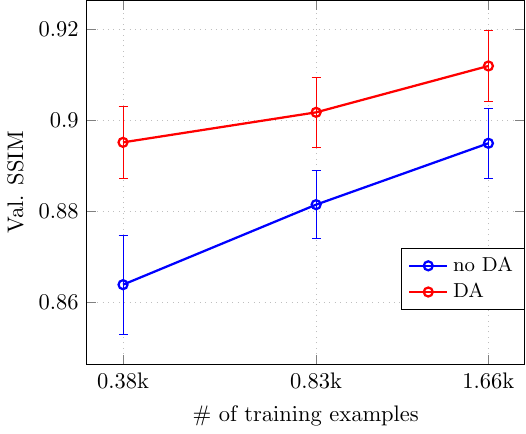}
		\caption{Validation SSIM vs. number of training images. Mean and standard error over $5$ train/val splits is depicted.}
		\label{fig:stanford2d_ssim_compare}
	\end{subfigure}\hfil 
	\caption{Experimental results on the Stanford 2D FSE dataset.	\vspace{-0.2cm}}
\end{figure}

\begin{figure}[htb]
	\captionsetup[subfigure]{labelformat=empty}
	\centering 
	\begin{subfigure}{0.23\textwidth}
		\includegraphics[width=\linewidth]{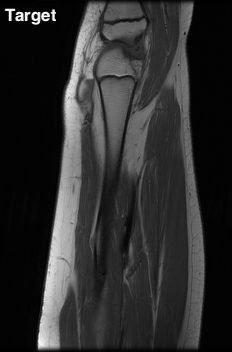}
		\vspace{-0.30cm}
	\end{subfigure}\hfil 
	\begin{subfigure}{0.23\textwidth}
		\includegraphics[width=\linewidth]{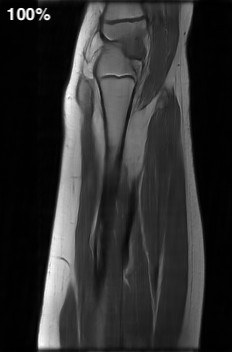}
		\vspace{-0.30cm}
	\end{subfigure}\hfil 
	\begin{subfigure}{0.23\textwidth}
		\includegraphics[width=\linewidth]{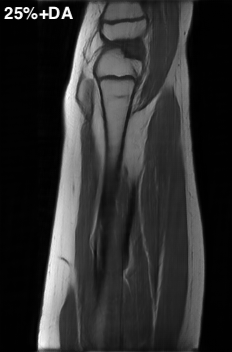}
		\vspace{-0.30cm}
	\end{subfigure}\hfil 
	\begin{subfigure}{0.23\textwidth}
		\includegraphics[width=\linewidth]{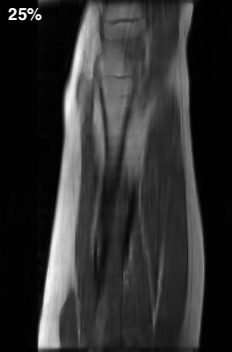}
		\vspace{-0.30cm}
	\end{subfigure}\hfil 
	\vspace{-0.2cm}
	\caption{Visual comparison of reconstructions on the Stanford 2D FSE dataset with and without data augmentation.\vspace{-0.3cm}}
	\label{fig:stanford2d_vis}
\end{figure}

We plot a training curve of validation SSIM with and without data augmentation in Figure \ref{fig:stanford2d_ssim_epochs}. The regularizing effect of data augmentation prevents overfitting to the training set and improves reconstruction SSIM on the validation dataset even in case of training $4\times$ longer than in the baseline experiments without data augmentation. Figure \ref{fig:stanford2d_ssim_compare} compares mean validation SSIM when the model is trained in different data regimes from $25\%$ to $100\%$ of all training data. MRAugment leads to significant improvement in reconstruction SSIM and this improvement is consistent across different train-val splits and training set sizes. We achieve higher mean SSIM using only $25\%$ of the training data with MRAugment than training on the full dataset without DA. On the full dataset, we improve reconstruction SSIM  from $0.8950$ to $0.9120$, and MRAugment achieves even larger gains in the lower data regime.  Figure \ref{fig:stanford2d_vis} provides a visual comparison of a reconstructed slice emphasizing the benefit of data augmentation. 

\textbf{Stanford Fullysampled 3D FSE Knees dataset.} The Stanford Fullysampled 3D FSE Knees dataset \citep{mridata:sawyer2013creation} consists of $20$ fully-sampled k-space volumes of knees. We use the same methodology to generate training and validation splits and evaluate results as in case of the Stanford 2D FSE dataset.

\begin{figure}[t]
	\centering
	\begin{subfigure}[t]{0.45\textwidth}
		\includegraphics[width=\linewidth]{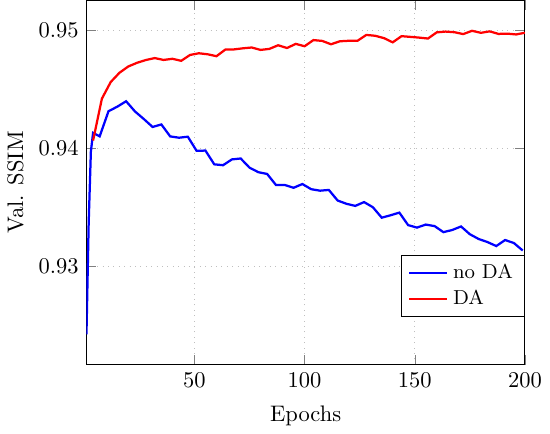}
		\caption{Validation SSIM vs. training epochs. We observe strong overfitting without data augmentation.}
		\label{fig:stanford3d_ssim_epochs}
	\end{subfigure}\hfil 
	\begin{subfigure}[t]{0.43\textwidth}
		\includegraphics[width=\linewidth]{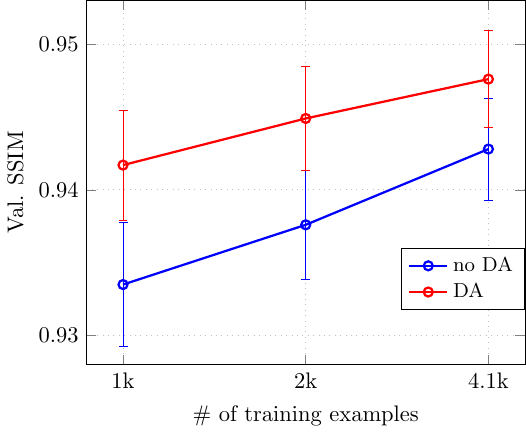}
		\caption{Validation SSIM vs. number of training images. Mean and standard error over $5$ train/val splits is depicted. }
		\label{fig:stanford3d_ssim_compare}
	\end{subfigure}\hfil 
	\caption{Experimental results on the Stanford Fullysampled 3D FSE dataset.}
\end{figure}

This dataset has significantly less variation compared to the Stanford 2D FSE dataset. Consequentially, we observe strong overfitting early in training if no data augmentation is used (Figure \ref{fig:stanford3d_ssim_epochs}). However, applying data augmentation successfully prevents overfitting. Furthermore, in accordance with observations on the Stanford 2D FSE dataset, data augmentation significantly boosts reconstruction SSIM across different data regimes (Figure \ref{fig:stanford3d_ssim_compare}). 

\subsection{High-data regime}
Next, we perform an extensive study on the fastMRI dataset \citep{fastmri:zbontar2018fastmri}, the largest publicly available fully-sampled MRI dataset with competitive baseline models, that allows us to investigate the utility of MRAugment across a wide range of training data regimes. More specifically, we use the fastMRI knee dataset, for which the original training set consists of approximately $35k$ MRI slices in $973$ volumes and we subsample to $1\%$, $10\%$, $33\%$ and $100\%$ of the original size. We measure performance on the original (fixed) validation set separate from the training set. 

\begin{figure}[t!]
	\captionsetup[subfigure]{labelformat=empty}
	\centering 
	\begin{subfigure}{0.23\textwidth}
		\includegraphics[width=\linewidth]{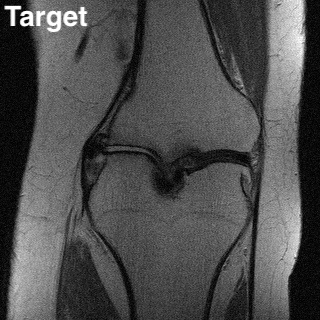}
		\label{fig:sc1_v2}
	\end{subfigure}\hfil 
	\begin{subfigure}{0.23\textwidth}
		\includegraphics[width=\linewidth]{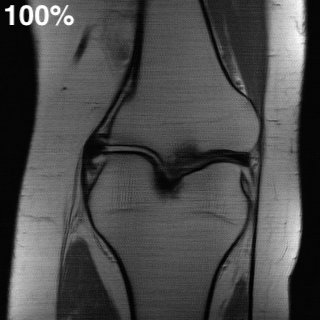}
		\label{fig:sc2_v2}
	\end{subfigure}\hfil 
	\begin{subfigure}{0.23\textwidth}
		\includegraphics[width=\linewidth]{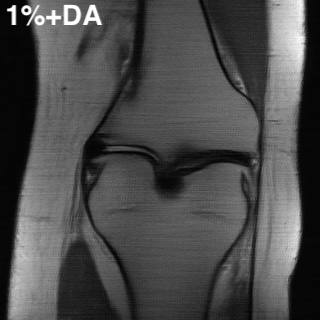}
		\label{fig:sc3_v2}
	\end{subfigure}\hfil 
	\begin{subfigure}{0.23\textwidth}
		\includegraphics[width=\linewidth]{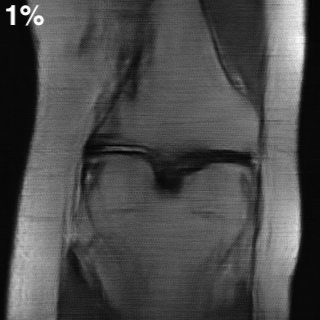}
		\label{fig:sc4_v2}
	\end{subfigure}
	\begin{subfigure}{0.23\textwidth}
		\includegraphics[width=\linewidth]{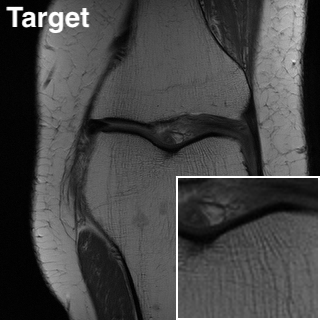}
		\label{fig:mc1_v2}
	\end{subfigure}\hfil 
	\begin{subfigure}{0.23\textwidth}
		\includegraphics[width=\linewidth]{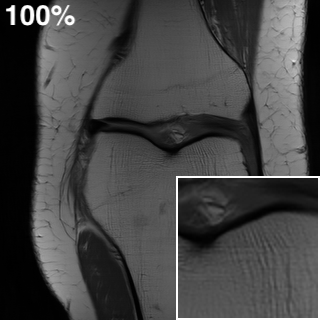}
		\label{fig:mc2_v2}
	\end{subfigure}\hfil 
	\begin{subfigure}{0.23\textwidth}
		\includegraphics[width=\linewidth]{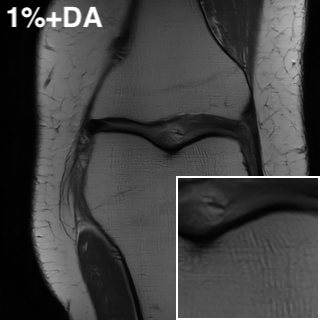}
		\label{fig:mc3_v2}
	\end{subfigure}\hfil 
	\begin{subfigure}{0.23\textwidth}
		\includegraphics[width=\linewidth]{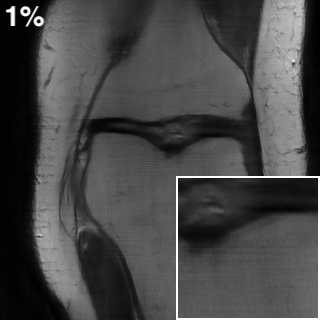}
		\label{fig:mc4_v2}
	\end{subfigure}\hfil 
	\vspace{-0.5cm}
	\caption{Visual comparison of single-coil (top row) and multi-coil (bottom-row) reconstructions using varying amounts of training data with and without data augmentation. We achieve reconstruction quality comparable to the state of the art but using 1\% of the training data. Without DA fine details are  completely lost.}
	\label{fig:combined_vis_v2}
\end{figure}
\begin{figure}[t!]
	\centering
	\begin{minipage}[b]{\textwidth}
		\centering
		\begin{subfigure}{\textwidth}
			\includegraphics[width=0.49\linewidth]{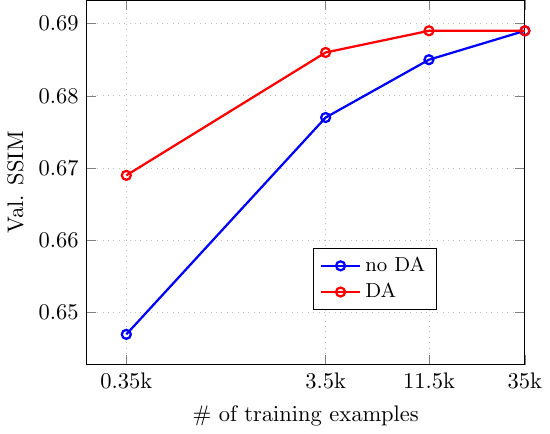}
			\hfill
			\includegraphics[width=0.49\linewidth]{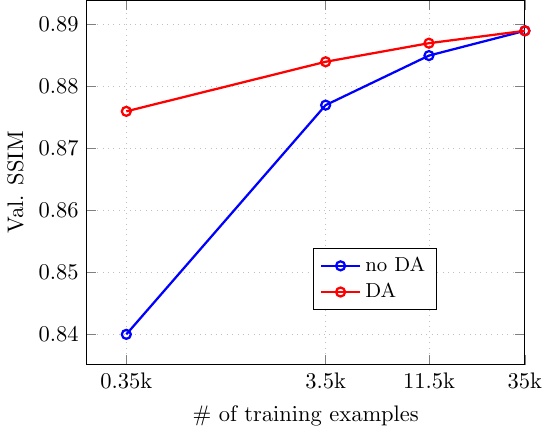}
		\end{subfigure}
		\caption{Single-coil (left) and multi-coil (right) validation SSIM vs. \# of training images. }
		\label{fig:ssim_compare}
	\end{minipage}%
\end{figure}

\textbf{Single-coil experiments.} For single-coil acquisition we are able to exactly match the performance of the model trained on the full dataset using only a third of the training data as depicted on the left in Fig. \ref{fig:ssim_compare}. Moreover, with only $10\%$ of the training data we achieve comparable SSIM to the model trained on the full dataset. The visual difference between reconstructions with and without data augmentation becomes striking in the low-data regime. As seen in the top row of Fig. \ref{fig:combined_vis_v2}, the model without DA was unable to reconstruct any of the fine details and the results appear blurry with strong artifacts. Applying MRAugment greatly improves reconstruction quality both in a quantitative and qualitative sense, visually approaching that obtained from training on the full dataset but using hundred times less data. 

\textbf{Multi-coil experiments.} As depicted on the right in Fig. \ref{fig:ssim_compare} for multi-coil acquisition we closely match the state of the art while significantly reducing training data. More specifically, we approach the state of the art SSIM within $0.6 \%$ using $10\%$ of the training data and within $0.25\%$ with $33\%$ of training data. As seen in the bottom row of Fig. \ref{fig:combined_vis_v2}, when using only $1\%$ of the training data we successfully reconstruct fine details comparable to that obtained from training on the full dataset, while high frequencies are completely lost without DA. 

Finally, we perform ablation studies on the fastMRI dataset and demonstrate that both pixel preserving and interpolating transformations individually improve reconstruction SSIM. Furthermore their effect is complementary: the best results are obtained by adding all transformations to the pipeline. Moreover, we investigate the effect of the data augmentation scheduling function and demonstrate that exponential scheduling results in better performance compared to a constant augmentation probability. We also evaluate a range of different augmentation schedules and show that both significantly lower or higher probabilities lead to poorer reconstruction SSIM. All ablation experiments can be found in the supplementary material.

\subsection{Model robustness\label{sec:robustness}}
In this section we investigate further potential benefits of data augmentation in scenarios where training examples from the target data distribution are not only scarce as studied before, but unavailable.
Distribution shifts can have a detrimental effect on a variety of reconstruction methods~\cite{Darestani_Chaudhari_Heckel_2021}.  Furthermore, we show some initial experimental results how data augmentation may help avoiding hallucinated features appearing on reconstructions due to overfitting.

\textbf{Unseen MR scanners.} First, we explore how data augmentation impacts generalization to new MRI scanner models not available in training time. Different MRI scanners may use different field strenghts for acquisition, and higher field strength typically correlates with higher SNR. Approximately half of the volumes in the fastMRI knee dataset have been acquired by a $1.5T$ scanner, whereas the rest by three different $3T$ scanners. We perform the following experiments:
\vspace{-0.2cm}
\begin{itemize}
	\item $3T \rightarrow 3T$: We train and validate on volumes acquired using $3T$ scanners. Volumes in the validation set have been imaged by a $3T$ scanner not in the training set. 
	\vspace{-0.1cm}
	\item $3T \rightarrow 1.5T$: We train on all volumes acquired by $3T$ scanners and validate on the $1.5T$ scanner. 
	\vspace{-0.1cm}
	\item $1.5T \rightarrow 3T$: We train on all volumes acquired by the $1.5T$ scanner and validate on all other $3T$ scanners. \vspace{-0.1cm}
\end{itemize}
Table \ref{tab:scanner_transfer} summarizes our results. Data augmentation consistently improves reconstruction SSIM on unseen scanner models. Similarly to our main experiments, the improvement is especially significant in the low-data regime. We observe that DA provides the greatest benefit when training on $1.5T$ scanners and testing on $3T$ models. We hypothesize that data augmentation can hinder the model from overfitting to the higher noise level present on $1.5T$ acquisitions during training thus resulting in better generalization on the lower noise $3T$ volumes.

\textbf{Unseen anatomies.} We demonstrate how data augmentation may help improving generalization on new anatomies not included in the training set, even in the high-data regime. We train a VarNet model on the complete fastMRI knee train dataset using the hyperparameters recommended in \citet{varnet:sriram2020end}, and evaluated the network on the fastMRI brain validation dataset throughout training. We repeated the experiment with the same hyperparameters, but with MRAugment turned on. The results can be seen in Fig. \ref{fig:transfer}. The regularizing effect of data augmentation impedes the network to overfit to the training dataset, thus the resulting model is more robust to shifts in test distribution. This results in higher reconstruction quality in terms of SSIM on unseen brain data when using MRAugment.

\begin{figure}
	\centering
	\begin{subfigure}{0.9\textwidth}
		\includegraphics[width=0.31\linewidth]{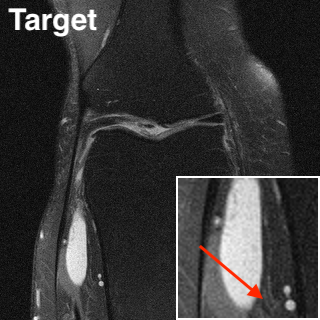}
		\includegraphics[width=0.31\linewidth]{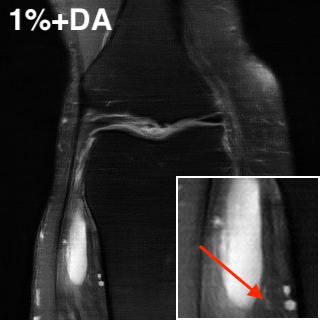}
		\includegraphics[width=0.31\linewidth]{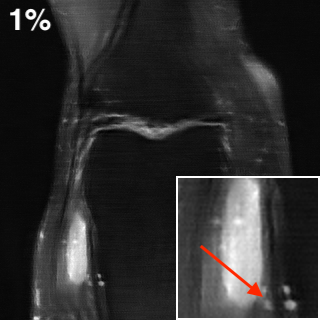} 	\vspace{0.1cm}
	\end{subfigure}
	\begin{subfigure}{0.9\textwidth}
		\includegraphics[width=0.31\linewidth]{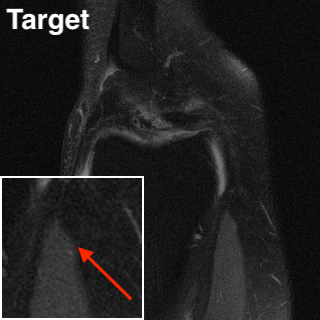}
		\includegraphics[width=0.31\linewidth]{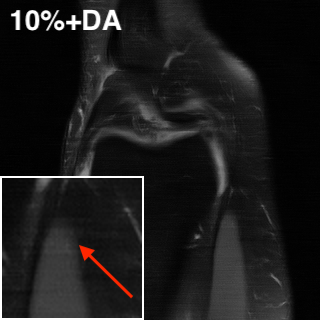}
		\includegraphics[width=0.31\linewidth]{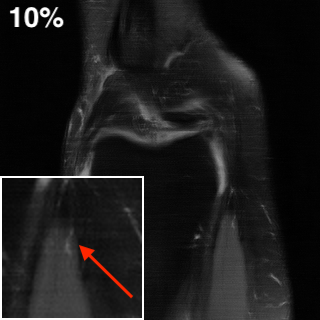}
	\end{subfigure}
	\caption{Hallucinated features appear on reconstructions without data augmentation.}
	\label{ap:fig:hallucination}
\end{figure}

\textbf{Hallucinations.} An unfortunate side-effect of deep learning based reconstructions may be the presence of hallucinated details. This is especially problematic in providing accurate medical diagnosis and lessens the trust of medical practicioners in deep learning. We observe that data augmentation has the potential benefit of increased robustness against hallucinations by preventing overfitting to training data, as seen in Fig. \ref{ap:fig:hallucination}. 

\begin{table}
	\begin{minipage}[c]{0.48\textwidth}
		\centering
		\resizebox{6.7cm}{!}{
			\begin{tabular}{|l|c|c|}
				\hline
				2\% train & no DA & DA\\ \hline
				$3T \rightarrow 3T$&  0.8646 & \textbf{0.9049} \\ 
				$3T \rightarrow 1.5T$& 0.8241 & \textbf{0.8551} \\ 
				$1.5T \rightarrow 3T$&  0.8174& \textbf{0.8913} \\  \hline
			\end{tabular} 
		}
		\resizebox{6.7cm}{!}{
			\begin{tabular}{|l|c|c|}
				\hline
				100\% train & no DA & DA\\ \hline
				$3T \rightarrow 3T$&  0.9177 & \textbf{0.9185} \\ 
				$3T \rightarrow 1.5T$& 0.8686 & \textbf{0.8690} \\ 
				$1.5T \rightarrow 3T$&  0.9043& \textbf{0.9062} \\  \hline
			\end{tabular}
		}
		\caption{Scanner transfer results using $2 \%$ (top) and $100 \%$ (bottom) of training data.  \label{tab:scanner_transfer}}
	\end{minipage}
	\hfill
	\begin{minipage}[c]{0.48\textwidth}
		\centering
		\includegraphics[width=0.85\linewidth]{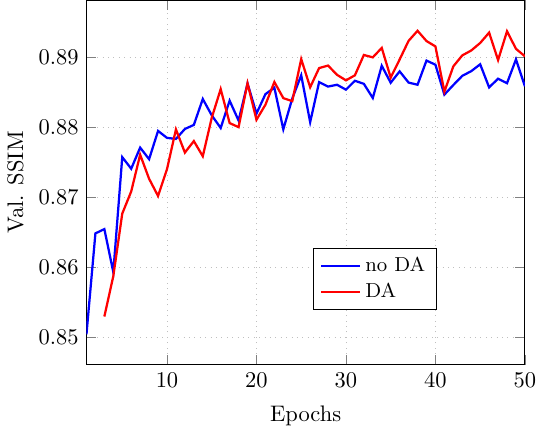}
		\vspace{-0.3cm}
		\captionof{figure}{Val. performance of models trained on knee and evaluated on brain MRI. \label{fig:transfer}}
	\end{minipage}
\end{table}

\subsection{Naive data augmentation \label{sec:exp_noise}} 
We would like to emphasize the importance of applying DA in a way that takes into account the measurement noise distribution. When applied incorrectly, DA leads to significantly worse performance than not using any augmentation. 

\begin{figure}[ht]
	\centering 
	\begin{subfigure}[b]{0.3\textwidth}
		\centering
		\includegraphics[width=0.96\linewidth]{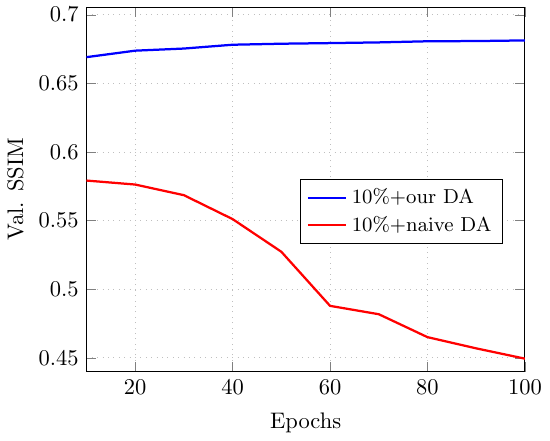}
		\caption{Naive data augmentation degrades generalization performance.}
		\label{fig:ssim_badDA_plot}
	\end{subfigure}
	\hfill
	\begin{subfigure}[b]{0.65\textwidth}
		\centering
		\includegraphics[width=0.32\linewidth]{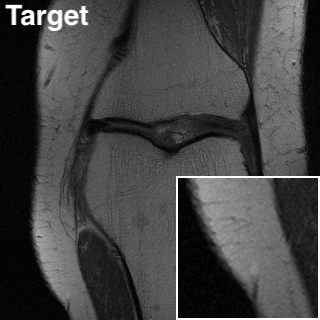}
		\includegraphics[width=0.32\linewidth]{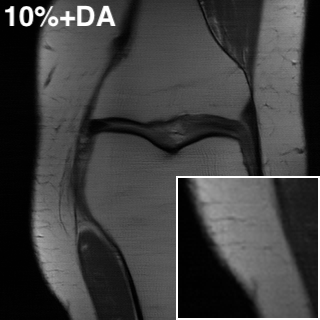}
		\includegraphics[width=0.32\linewidth]{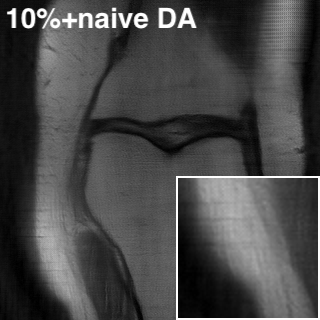}
		\vspace{0.3cm}
		\caption{ Naive data augmentation that does not preserve measurement noise distribution leads to significantly degraded reconstruction quality. }
		\label{fig:badDA_vis}
	\end{subfigure} 
	\caption{Experimental results comparing MRAugment with naive data augmentation.}
	\label{fig:bad_da_compare}
\end{figure}

We train a model using 'naive' data augmentation without considering the measurement noise distribution as described in Section \ref{sec:noise}, by augmenting real-valued target images. We use the same exponential augmentation probability scheduling for MRAugment and the naive approach. As Fig. \ref{fig:ssim_badDA_plot} demonstrates, reconstruction quality degrades over training using the naive approach. This is due to the fact that as augmentation probability increases, the network sees less and less original, unaugmented images, whereas the poorly augmented images are detrimental for generalization due to the mismatch in train and validation noise distribution. On the other hand, MRAugment clearly helps and validation performance steadily improves over epochs. Fig. \ref{fig:badDA_vis} provides a visual comparison of reconstructions using naive DA and our data augmentation method tailored to the problem. Naive DA reconstruction exhibits high-frequency artifacts and low image quality caused by the mismatch in noise distribution. These drastic differences underline the vital importance of taking a careful, physics-informed approach to data augmentation for MR reconstruction.

\section{Conclusion}
In this paper, we develop a physics-based data augmentation pipeline for accelerated MR imaging. We find that MRAugment yields significant gains in the low-data regime which can be beneficial in applications where only little training data is available or where the training data changes quickly. We also demonstrate that models trained with data augmentation are more robust against overfitting and shifts in the test distribution.
We believe that this work opens up many interesting venues for further research with respect to data augmentation for inverse problems in the low data regime. First, learning efficient data augmentation from the training data has been investigated in prior literature \citep{da:cubuk2019autoaugment, da:lemley2017smart, da:tran2017bayesian} and would be a natural extension of our method. Second, finding the optimal augmentation strength throughout training is challenging, therefore an adaptive scheme that automatically schedules the augmentation probability would potentially further improve upon our results. Such a method is proposed in \citet{ganda:karras2020training} for discriminator augmentation in GAN training, where the augmentation strength is adjusted based on an discriminator overfitting heuristic with great success. Finally, combining our technique with a generative framework such as AmbientGAN \citep{ganda:bora2018ambientgan} that generates high quality samples of a target distribution from noisy partial measurements, could be potentially used to synthesize fully-sampled MRI data from few noisy k-space measurements.

\section*{Acknowledgments}
M. Soltanolkotabi is supported by the Packard Fellowship in Science and Engineering, a Sloan Research Fellowship in Mathematics, an NSF-CAREER under award \#1846369, DARPA Learning with Less Labels (LwLL) and FastNICS programs, and NSF-CIF awards \#1813877 and \#2008443.
R.~Heckel is supported by the IAS at TUM, the DFG (German Research Foundation), and by the NSF under award IIS-1816986. 

\bibliography{FastMRIDA}
\bibliographystyle{plainnat}

\newpage
\appendix
\appendix
\section*{Appendix outline}
The following appendix provides additional experimental details, enlarged images of reconstructed slices and extra discussions not included in the main paper. The organization of the supplementary material is as follows:

\textbf{FastMRI dataset.} Appendix \ref{appendix:fastMRI} provides additional details on the fastMRI dataset and the experimental setup used in our experiments. We plot reconstruction metric results in PSNR in addition to the SSIM comparison in the main paper (Fig. \ref{fig:ssim_compare}) and demonstrate gains comparable to that measured in SSIM.  Furthermore, we plot randomly picked reconstructions from the validation set in order to provide a comprehensive view of reconstruction quality. Finally, we apply MRAugment with a model different from the one used in our main experiments on the fastMRI dataset to demonstrate the wider applicability of DA for deep learning based MR reconstruction. 

\textbf{Stanford datasets.} In Appendices \ref{appendix:stanford2d} and \ref{appendix:stanford3d} we provide more details on the Stanford datasets and the experimental details. Moreover, further reconstructed slices are depicted complementing the ones in the main paper. 

\textbf{Robustness.} In Appendix \ref{appendix:robustness} we give more details on the robustness experiments from Section \ref{sec:robustness}, describing the MR scanner models used in the experiments.

\textbf{Ablation studies.} In Appendix \ref{appendix:ablation} we perform ablation studies on the fastMRI dataset to investigate the utility of various augmentations and the effect of augmentation scheduling on the final reconstruction. 


Finally, our code is published at \url{https://github.com/MathFLDS/MRAugment}. We refer to this code for additional detail regarding the implementation. We note that MRAugment pipeline can be seamlessly integrated with any existing MR reconstruction code, and can be applied to the fastMRI code base by only a couple of lines of additional code. We hope that the utility and ease of use of MRAugment will prove useful for a wider range of practitioners.

\section{Experiments on the fastMRI dataset \label{appendix:fastMRI}}
\subsection{Experimental details}
The fastMRI dataset \citep{fastmri:zbontar2018fastmri} is a large open dataset of knee and brain MRI volumes. The train and validation splits contain fully-sampled k-space volumes and corresponding target reconstructions for both (simulated) single-coil and multi-coil acquisition. The knee MRI dataset we are focusing on in this paper includes $973$ train volumes ($34742$ slices) and $199$ validation volumes ($7135$ slices). The target reconstructions are fixed size $320 \times 320$ center cropped images corresponding to the fully-sampled data of varying sizes. The undersampling ratio is either $25\%$ ($4\times$ acceleration) or $12.5\%$ ($8\times$ acceleration). Undersampling is performed along the phase encoding dimension in k-space, that is columns in k-space are sampled.  A certain neighborhood of adjacent low-frequency lines are always included in the measurement. The size of this fully-sampled region is $8\%$ of all frequencies in case of $4\times$ acceleration and $4\%$ in case of $8\times$ acceleration. 


\textbf{Dataset sampling.} We use the fastMRI \citep{fastmri:zbontar2018fastmri} single-coil and multi-coil knee dataset for our experiments. For creating the sub-sampled datasets, we uniformly sample volumes from the training set, and add all slices from the sampled volumes. Our validation results are reported on the whole validation dataset. Images in the dataset have varying dimensions. Due to GPU memory considerations we center-cropped the input images to $640 \times 368$ pixels (which covers most of the images). We use random undersampling masks with $8\times$ acceleration and $4\%$ fully-sampled low-frequency band, undersampled in the phase encoding direction by masking whole kspace lines. We generate a new random mask for each slice on-the-fly while training, but use the same fixed mask for each slice within the same volume on the validation set (different across volumes).  

\textbf{Model.} We train the default E2E-VarNet network from \citet{varnet:sriram2020end} with $12$ cascades (approx. $30M$ parameters) for both the single-coil and multi-coil reconstruction problems. For single-coil data we remove the Sensitivity Map Estimation sub-network as sensitivity maps are not relevant in this problem.  

\textbf{Hyperparameters and training.} We use an Adam optimizer with $0.0003$ learning rate following \citet{varnet:sriram2020end}. We train the baseline model on the full training dataset for $50$ epochs. For the smaller, sub-sampled datasets we train for the same computational cost as the baseline, that is we train for $N \cdot 50$ epochs on $1/N$th of the training data. Without data augmentation, we observe a saturation in validation SSIM during this time. With data augmentation we trained $50\%$ longer as we still observe improvement in validation performance after the standard number of epochs. We report the best SSIM on the validation set throughout training. We train on 4 GPUs for single-coil data and on 8 GPUs for multi-coil data. The batch size matches the number of GPUs used for training, since a GPU can only hold a single datapoint.

\textbf{Data augmentation parameters.} The transformations and their corresponding probability weights and ranges of values are depicted in Table \ref{tab:da_params}. We adjust the weights so that groups of transformations such as rotation (arbitrary, by $k \cdot 90^\circ$), flipping (horizontal or vertical) or scaling (isotropic or anisotropic) have similar probabilities. For both the affine transformations and upsampling we use bicubic interpolation. Due to computational considerations we only use upsampling before transformations for the single-coil experiments.

\subsection{Additional experimental results on the fastMRI dataset }
\begin{figure}[h!]
	\centering
	\begin{subfigure}{0.96\textwidth}
		\includegraphics[width=0.49\linewidth]{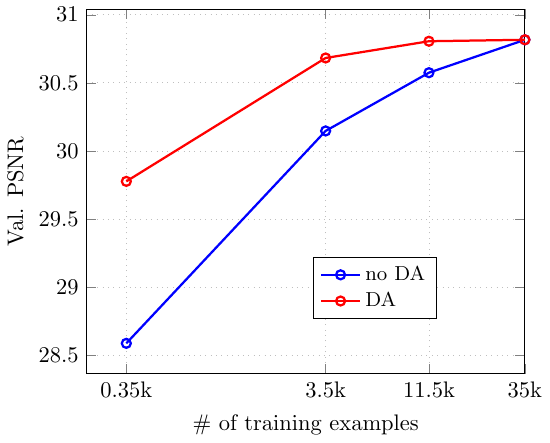}
		\hfill
		\includegraphics[width=0.48\linewidth]{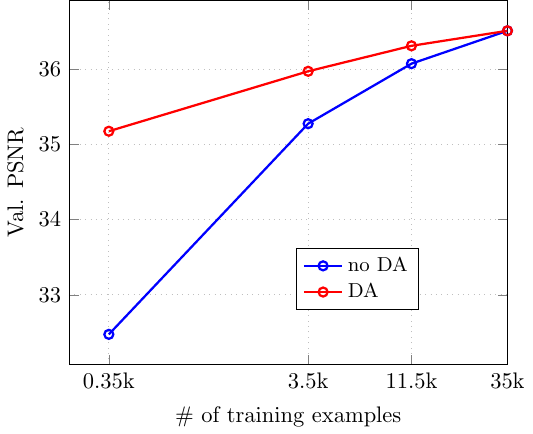}
	\end{subfigure}
	\caption{Single-coil (left) and multi-coil (right) validation PSNR vs. \# of training images. }
	\label{fig:psnr_compare}
\end{figure}

\textbf{Comparison of additional metrics.}
In order to provide more in-depth comparison for our main experiment, here we provide results on PSNR as an additional image quality metric, extending our results from Figure \ref{fig:ssim_compare}. We observe significant and consistent improvement in PSNR when applying MRAugment (Fig. \ref{fig:psnr_compare}) with similar trends to SSIM: the improvement is the most prominent in the low-data regime, but still significant in the moderate domain. 

\textbf{Additional reconstructions.} In order to demonstrate that MRAugment works well across a wide range of MR slices, here we provide additional reconstructions with and without data augmentation.  In multi-coil reconstructions the visual differences are more subtle, therefore we magnified regions with fine details for better comparison.

\begin{figure}
	\captionsetup[subfigure]{labelformat=empty}
	\centering 
	\begin{subfigure}{0.24\textwidth}
		\includegraphics[width=\linewidth]{plots/recon_compare/singlecoil/recon_single_train100.png}
		\label{fig:sc1}
		\vspace{-0.30cm}
	\end{subfigure}\hfil 
	\begin{subfigure}{0.24\textwidth}
		\includegraphics[width=\linewidth]{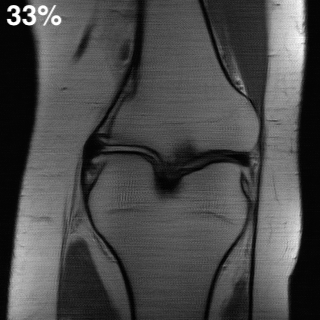}
		\label{fig:sc2}
		\vspace{-0.30cm}
	\end{subfigure}\hfil 
	\begin{subfigure}{0.24\textwidth}
		\includegraphics[width=\linewidth]{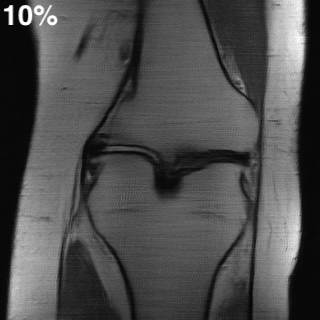}
		\label{fig:sc3}
		\vspace{-0.30cm}
	\end{subfigure}\hfil 
	\begin{subfigure}{0.24\textwidth}
		\includegraphics[width=\linewidth]{plots/recon_compare/singlecoil/recon_single_train1.png}
		\label{fig:sc4}
		\vspace{-0.30cm}
	\end{subfigure}\hfil 
	\begin{subfigure}{0.24\textwidth}
		\includegraphics[width=\linewidth]{plots/recon_compare/singlecoil/target.png}
		\label{fig:sc6}
		\vspace{-0.30cm}
	\end{subfigure}\hfil 
	\begin{subfigure}{0.24\textwidth}
		\includegraphics[width=\linewidth]{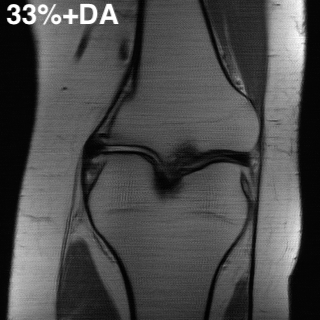}
		\label{fig:sc7}
		\vspace{-0.30cm}
	\end{subfigure}\hfil 
	\begin{subfigure}{0.24\textwidth}
		\includegraphics[width=\linewidth]{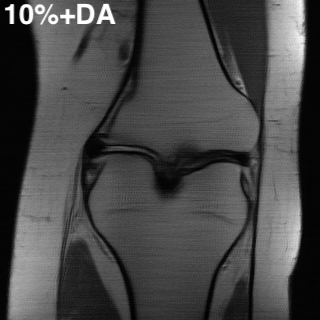}
		\label{fig:sc8}
		\vspace{-0.30cm}
	\end{subfigure}\hfil 
	\begin{subfigure}{0.24\textwidth}
		\includegraphics[width=\linewidth]{plots/recon_compare/singlecoil/recon_single_train1_DA.png}
		\label{fig:sc9}
		\vspace{-0.30cm}
	\end{subfigure}
	\caption{Visual comparison of fastMRI single-coil reconstructions presented in Figure \ref{fig:combined_vis_v2} extended with additional images corresponding to various amount of training data.}
	\label{fig:singlecoil_vis}
\end{figure}

\begin{figure}
	\captionsetup[subfigure]{labelformat=empty}
	\centering 
	\begin{subfigure}{0.24\textwidth}
		\includegraphics[width=\linewidth]{plots/recon_compare/multicoil/recon_multi_train100.png}
		\label{fig:mc1}
		\vspace{-0.30cm}
	\end{subfigure}\hfil 
	\begin{subfigure}{0.24\textwidth}
		\includegraphics[width=\linewidth]{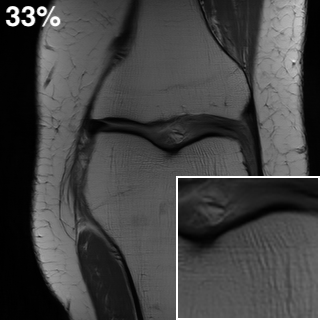}
		\label{fig:mc2}
		\vspace{-0.30cm}
	\end{subfigure}\hfil 
	\begin{subfigure}{0.24\textwidth}
		\includegraphics[width=\linewidth]{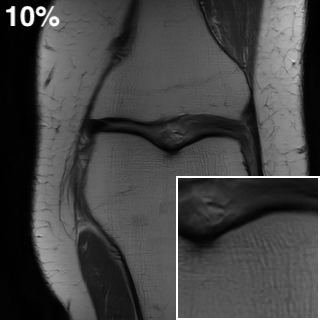}
		\label{fig:mc3}
		\vspace{-0.30cm}
	\end{subfigure}\hfil 
	\begin{subfigure}{0.24\textwidth}
		\includegraphics[width=\linewidth]{plots/recon_compare/multicoil/recon_multi_train1.png}
		\label{fig:mc4}
		\vspace{-0.30cm}
	\end{subfigure}\hfil 
	\begin{subfigure}{0.24\textwidth}
		\includegraphics[width=\linewidth]{plots/recon_compare/multicoil/target.png}
		\label{fig:mc6}
		\vspace{-0.30cm}
	\end{subfigure}\hfil 
	\begin{subfigure}{0.24\textwidth}
		\includegraphics[width=\linewidth]{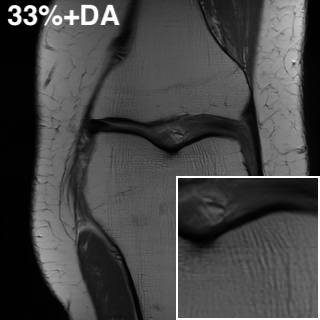}
		\label{fig:mc7}
		\vspace{-0.30cm}
	\end{subfigure}\hfil 
	\begin{subfigure}{0.24\textwidth}
		\includegraphics[width=\linewidth]{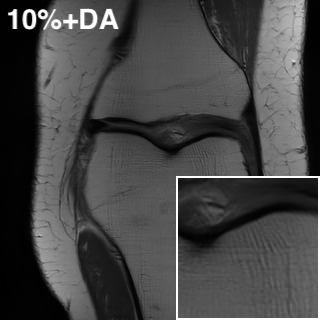}
		\label{fig:mc8}
		\vspace{-0.30cm}
	\end{subfigure}\hfil 
	\begin{subfigure}{0.24\textwidth}
		\includegraphics[width=\linewidth]{plots/recon_compare/multicoil/recon_multi_train1_DA.png}
		\label{fig:mc9}
		\vspace{-0.30cm}
	\end{subfigure}
	\caption{Visual comparison of fastMRI multi-coil reconstructions presented in Figure \ref{fig:combined_vis_v2} extended with additional images corresponding to various amount of training data.}
	\label{fig:multicoil_vis}
\end{figure}

Figures \ref{fig:singlecoil_vis} and \ref{fig:multicoil_vis} provide a comprehensive set of reconstructions across all subsampling ratios with and without data augmentation for the single-coil and multi-coil slices additional to the ones presented in Figure \ref{fig:combined_vis_v2}.

\begin{figure}[htb]
	\centering
	\begin{subfigure}[t]{0.97\textwidth}
		\includegraphics[width=0.325\textwidth]{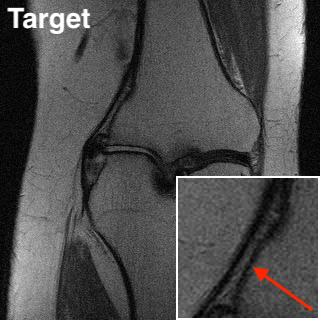}
		\hfill
		\includegraphics[width=0.325\textwidth]{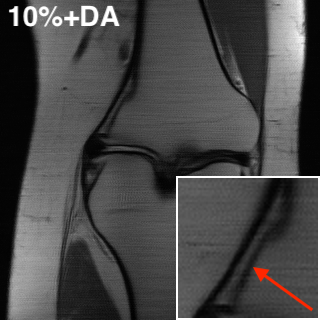}
		\hfill
		\includegraphics[width=0.325\textwidth]{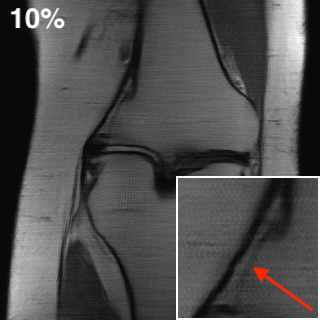}
	\end{subfigure}
	\caption{MRAugment recovers additional details in the moderate-data regime when $10\%$ of fastMRI training data is used.}
	\label{fig:midcompare}
\end{figure}

Even though the most visible improvement on reconstructions is observed when training data is especially low ($1 \%$ subsampling), Figure \ref{fig:midcompare} provides a closer look at a slice where significant details are recovered by MRAugment using $10 \%$ of training data.

\begin{figure}[htb]
	\captionsetup[subfigure]{labelformat=empty}
	\centering 
	\begin{subfigure}{0.242\textwidth}
		\caption{Ground truth}
		\includegraphics[width=\linewidth]{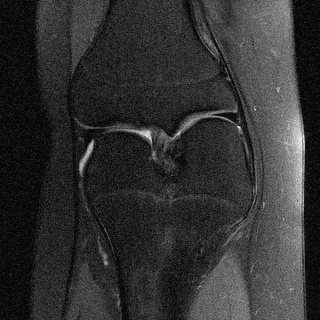}
	\end{subfigure}\hfil 
	\begin{subfigure}{0.242\textwidth}
		\caption{$100\%$ train}
		\includegraphics[width=\linewidth]{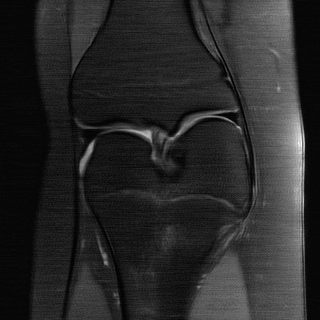}
	\end{subfigure}\hfil 
	\begin{subfigure}{0.242\textwidth}
		\caption{$1\%$ train + DA}
		\includegraphics[width=\linewidth]{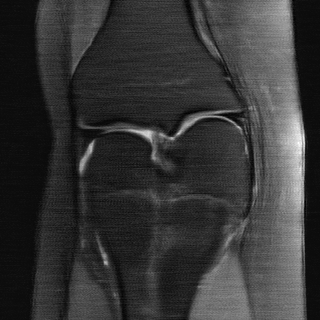}
	\end{subfigure}\hfil 
	\begin{subfigure}{0.242\textwidth}
		\caption{$1\%$ train}
		\includegraphics[width=\linewidth]{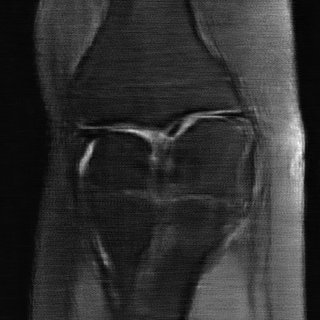}
	\end{subfigure}
	
	\begin{subfigure}{0.242\textwidth}
		\includegraphics[width=\linewidth]{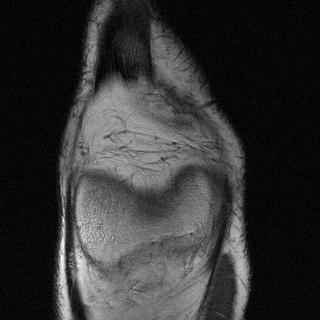}
	\end{subfigure}\hfil 
	\begin{subfigure}{0.242\textwidth}
		\includegraphics[width=\linewidth]{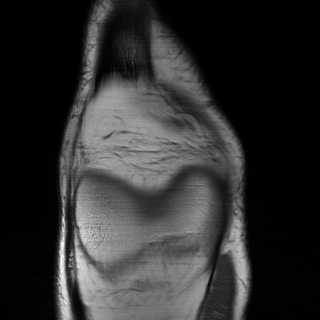}
	\end{subfigure}\hfil 
	\begin{subfigure}{0.242\textwidth}
		\includegraphics[width=\linewidth]{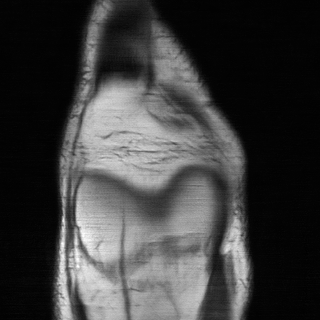}
	\end{subfigure}\hfil 
	\begin{subfigure}{0.242\textwidth}
		\includegraphics[width=\linewidth]{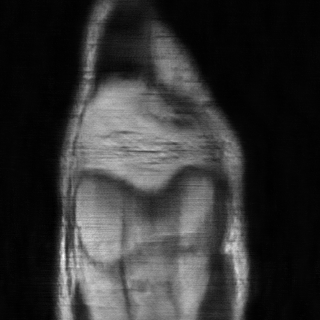}
	\end{subfigure}
	
	\begin{subfigure}{0.242\textwidth}
		\includegraphics[width=\linewidth]{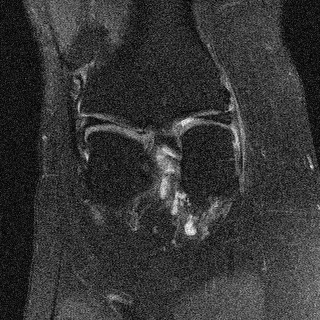}
	\end{subfigure}\hfil 
	\begin{subfigure}{0.242\textwidth}
		\includegraphics[width=\linewidth]{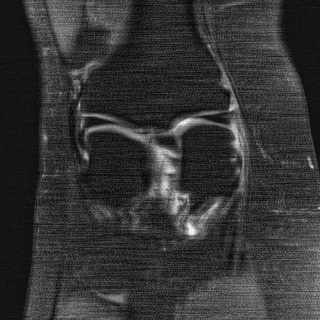}
	\end{subfigure}\hfil 
	\begin{subfigure}{0.242\textwidth}
		\includegraphics[width=\linewidth]{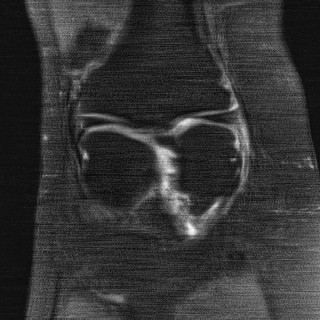}
	\end{subfigure}\hfil 
	\begin{subfigure}{0.242\textwidth}
		\includegraphics[width=\linewidth]{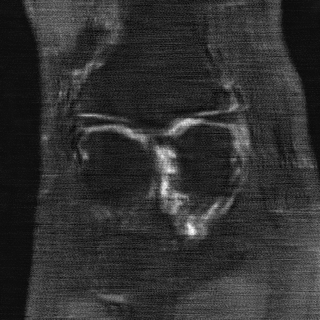}
	\end{subfigure}
	
	\begin{subfigure}{0.242\textwidth}
		\includegraphics[width=\linewidth]{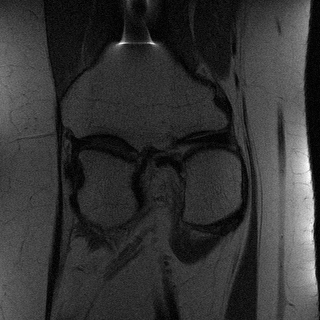}
	\end{subfigure}\hfil 
	\begin{subfigure}{0.242\textwidth}
		\includegraphics[width=\linewidth]{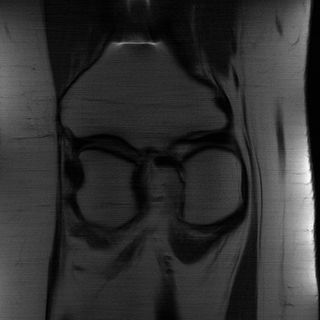}
	\end{subfigure}\hfil 
	\begin{subfigure}{0.242\textwidth}
		\includegraphics[width=\linewidth]{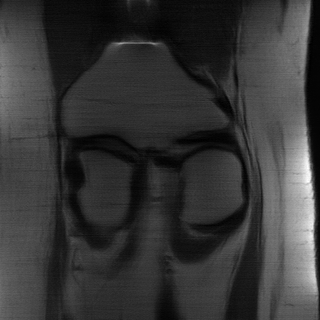}
	\end{subfigure}\hfil 
	\begin{subfigure}{0.242\textwidth}
		\includegraphics[width=\linewidth]{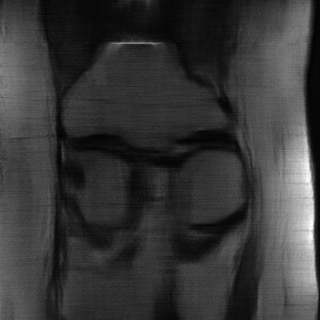}
	\end{subfigure}
	
	\begin{subfigure}{0.242\textwidth}
		\includegraphics[width=\linewidth]{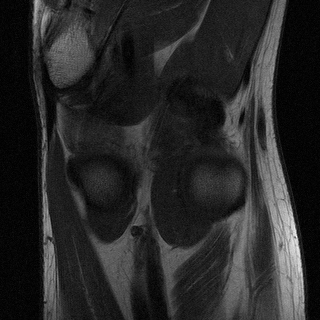}
	\end{subfigure}\hfil 
	\begin{subfigure}{0.242\textwidth}
		\includegraphics[width=\linewidth]{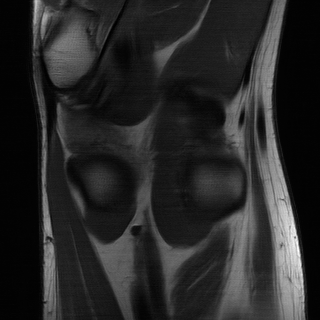}
	\end{subfigure}\hfil 
	\begin{subfigure}{0.242\textwidth}
		\includegraphics[width=\linewidth]{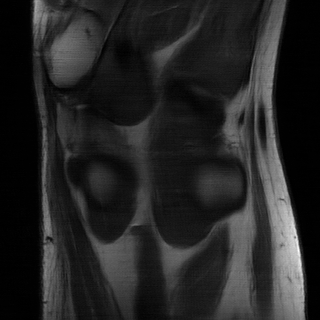}
	\end{subfigure}\hfil 
	\begin{subfigure}{0.242\textwidth}
		\includegraphics[width=\linewidth]{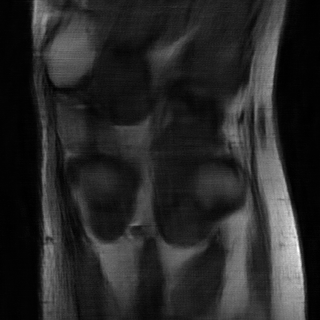}
	\end{subfigure}
	
	\caption{Visual comparison of fastMRI single-coil reconstructions using varying amounts of training data with and without data augmentation.}
	\label{fig:singlecoil_appendix}
\end{figure}

\begin{figure}[htb]
	\captionsetup[subfigure]{labelformat=empty}
	\centering 
	\begin{subfigure}{0.242\textwidth}
		\caption{Ground truth}
		\includegraphics[width=\linewidth]{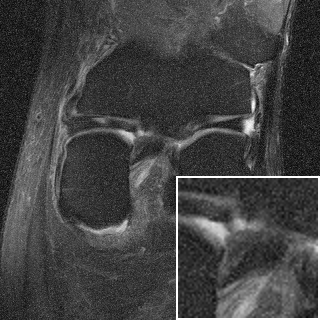}
	\end{subfigure}\hfil 
	\begin{subfigure}{0.242\textwidth}
		\caption{$100\%$ train}
		\includegraphics[width=\linewidth]{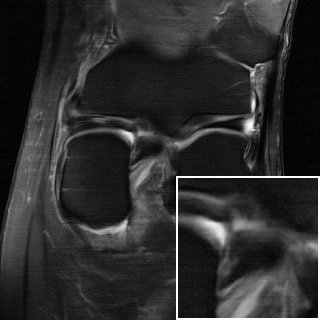}
	\end{subfigure}\hfil 
	\begin{subfigure}{0.242\textwidth}
		\caption{$1\%$ training data + DA}
		\includegraphics[width=\linewidth]{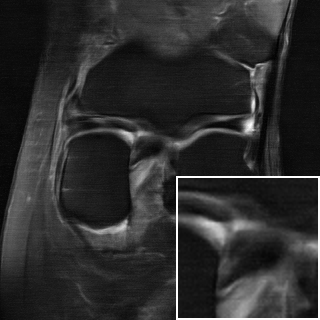}
	\end{subfigure}\hfil 
	\begin{subfigure}{0.242\textwidth}
		\caption{$1\%$ training data}
		\includegraphics[width=\linewidth]{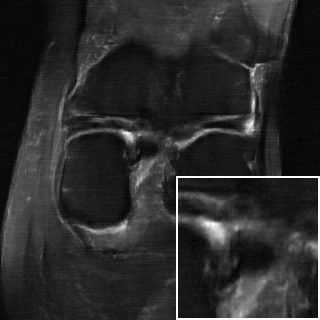}
	\end{subfigure}
	
	\begin{subfigure}{0.242\textwidth}
		\includegraphics[width=\linewidth]{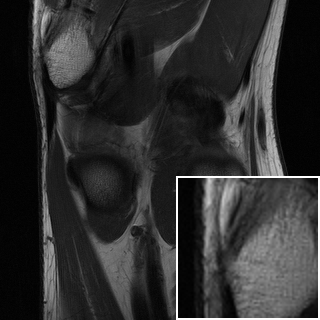}
	\end{subfigure}\hfil 
	\begin{subfigure}{0.242\textwidth}
		\includegraphics[width=\linewidth]{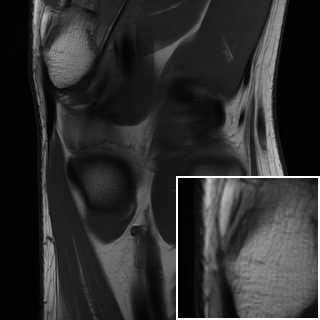}
	\end{subfigure}\hfil 
	\begin{subfigure}{0.242\textwidth}
		\includegraphics[width=\linewidth]{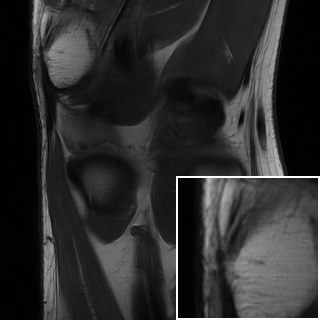}
	\end{subfigure}\hfil 
	\begin{subfigure}{0.242\textwidth}
		\includegraphics[width=\linewidth]{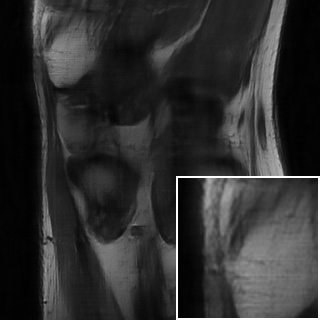}
	\end{subfigure}
	
	\begin{subfigure}{0.242\textwidth}
		\includegraphics[width=\linewidth]{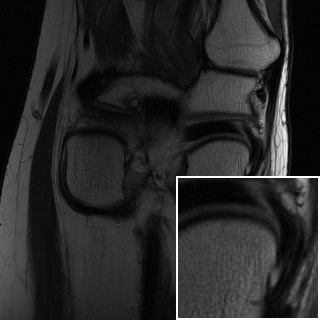}
	\end{subfigure}\hfil 
	\begin{subfigure}{0.242\textwidth}
		\includegraphics[width=\linewidth]{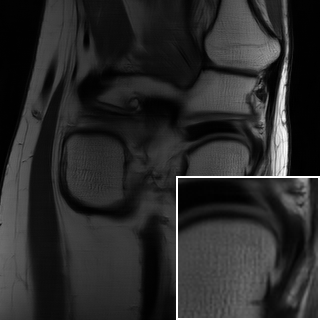}
	\end{subfigure}\hfil 
	\begin{subfigure}{0.242\textwidth}
		\includegraphics[width=\linewidth]{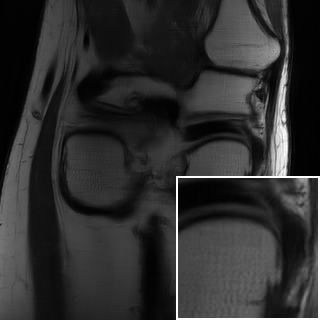}
	\end{subfigure}\hfil 
	\begin{subfigure}{0.242\textwidth}
		\includegraphics[width=\linewidth]{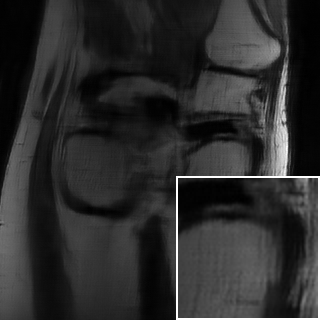}
	\end{subfigure}\hfil 
	
	\begin{subfigure}{0.242\textwidth}
		\includegraphics[width=\linewidth]{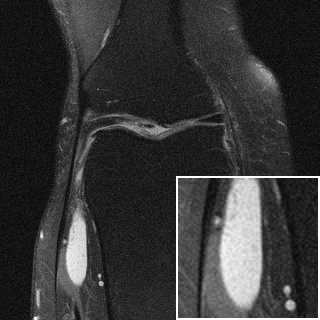}
	\end{subfigure}\hfil 
	\begin{subfigure}{0.242\textwidth}
		\includegraphics[width=\linewidth]{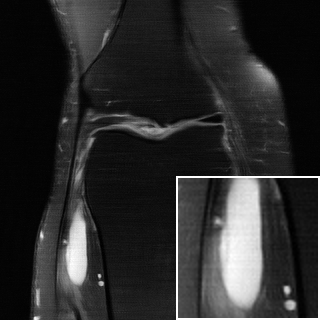}
	\end{subfigure}\hfil 
	\begin{subfigure}{0.242\textwidth}
		\includegraphics[width=\linewidth]{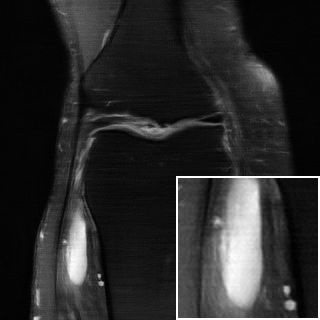}
	\end{subfigure}\hfil 
	\begin{subfigure}{0.242\textwidth}
		\includegraphics[width=\linewidth]{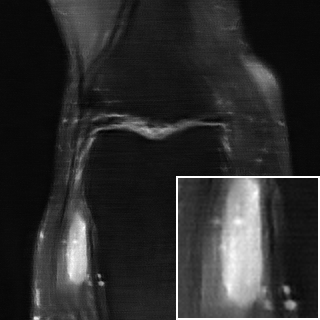}
	\end{subfigure}\hfil 
	
	\begin{subfigure}{0.242\textwidth}
		\includegraphics[width=\linewidth]{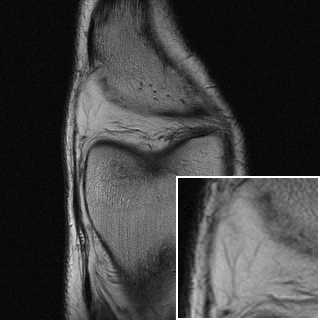}
	\end{subfigure}\hfil 
	\begin{subfigure}{0.242\textwidth}
		\includegraphics[width=\linewidth]{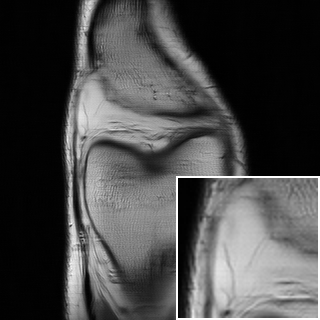}
	\end{subfigure}\hfil 
	\begin{subfigure}{0.242\textwidth}
		\includegraphics[width=\linewidth]{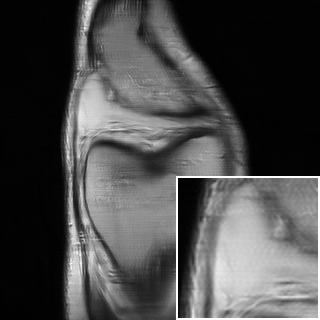}
	\end{subfigure}\hfil 
	\begin{subfigure}{0.242\textwidth}
		\includegraphics[width=\linewidth]{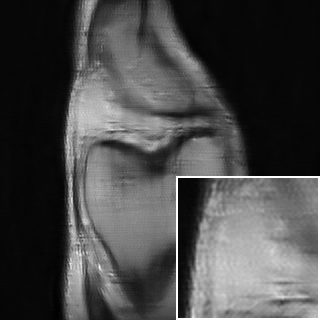}
	\end{subfigure}\hfil 
	
	\caption{Visual comparison of fastMRI multi-coil reconstructions using varying amounts of training data with and without data augmentation.}
	\label{fig:multicoil_appendix}
\end{figure}

Figures \ref{fig:singlecoil_appendix} and \ref{fig:multicoil_appendix} provide more reconstructed slices randomly sampled from the validation dataset with and without DA.

\textbf{Other models.} Even though we demonstrated our DA pipeline on E2E-VarNet, the potential of our technique is not limited to a specific model. We performed preliminary experiments on i-RIM \cite{irim:putzky2019invert}, another high performing model on single-coil MR reconstruction. We kept the hyperparameters proposed in \citet{irim:putzky2019invert} for the single-coil problem with modifications as follows. Due to computational considerations, we decreased the number of invertible layers to $6$ with $[64, 128, 256, 256, 128, 64]$ hidden features inside the reversible blocks and $[1, 2, 4, 4, 2, 1]$ kernel strides in each layer, resulting in a model with $20M$ parameters. In order to further reduce training time, we trained on volumes without fat suppression that take up $50\%$ of the full fastMRI knee dataset. We refer to this new reduced dataset as 'full' in this section. Finally, we trained on $368x368$ center crops of input images for each experiment. We used ramp scheduling with augmentation probability linearly increasing from $0$ to $p_{max} = 0.4$. The acceleration factor and undersampling mask were the same as for other experiments. As depicted in Fig. \ref{fig:irim} , our experiments show that applying data augmentation to only $10\%$ of the training data can match the performance of the model trained on the full dataset.


\begin{center}
	\includegraphics[width=0.5\linewidth]{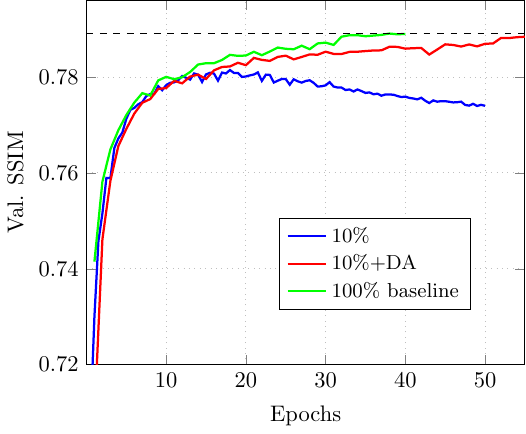}
	\captionof{figure}{Results on the i-RIM network. We achieve SSIM comparable to the baseline with only $10\%$ of the training data.}
	\label{fig:irim}
\end{center}


\section{Experiments on the Stanford 2D FSE dataset  \label{appendix:stanford2d}}
The Stanford 2D FSE \cite{website:Stanford2D} dataset is a public dataset of $89$ fully-sampled MRI volumes of various anatomies including lower extremity, pelvis and cardiac images. All measurements have been acquired by the same MRI scanner using multi-coil acquisition, however volume dimensions and the number of receiver coils vary between volumes. The total number of MRI slices is about $5 \%$ of the fastMRI knee training dataset.

\textbf{Dataset sampling.} When random sampling, we randomly select volumes of the original dataset and add all slices of the sampled volumes. For volumes where multiple contrasts are available, we arbitrarily pick the first one and discard the others. We scale all measurements by $10^{-7}$ to approximately match the range of fastMRI measurements. Unlike the fastMRI dataset, Stanford 2D FSE is not separated into training, validation and test sets. Therefore, we use $80\% - 20\%$ training-validation split in our experiments, where we generate $5$ random splits in order to minimize variations in reconstruction metrics due to validation set selection and show the mean of validation SSIMs over all $5$ runs. When performing experiments on less training data, we keep $20\%$ of the full dataset as validation set and only subsample the train split. We use no center-cropping on the training images as volume dimensions vary strongly. We undersample the measurements by a factor of $8$ and generate masks the same way as in the fastMRI experiments detailed in Section \ref{appendix:fastMRI}.  

\textbf{Model.} We train the default E2E-VarNet network as used in the multi-coil fastMRI experiments detailed in Section \ref{appendix:fastMRI}.

\textbf{Hyperparameters and training.} We use Adam optimizer with a learning rate of $0.0003$ as in our other experiments. For the baseline experiments without data augmentation, we train the model for $50$ epochs, after which we see no significant improvement in reconstruction SSIM and the model overfits to the training dataset. With data augmentation we train for $200$ epochs, as validation SSIM increases well after $50$ epochs and we observe no overfitting. In all experiments, we report the mean of best validation SSIMs over $5$ independent runs. 

\textbf{Data augmentation parameters.} In all data augmentation experiments on the Stanford 2D FSE dataset we use exponential schedulig with $p_{max} = 0.55$. The range of values for the various transformations is almost identical to that in Table \ref{tab:da_params}. For more details we refer the reader to the attached source code.

\section{Experiments on the Stanford Fullysampled 3D FSE Knees dataset \label{appendix:stanford3d}}
The Stanford Fullysampled 3D FSE Knees dataset \cite{mridata:sawyer2013creation} is a public MRI dataset of $20$ fully-sampled k-space volumes of knees, acquired by the same MRI scanner. Each volume consists of $256$ slices of $320 \times 320$ images with a multi-coil acquisition using $8$ receiver coils. The full dataset consists of $5120$ slices, or about $15 \%$ of the fastMRI knee training dataset.

\textbf{Experimental setup.} We apply the same dataset sampling, metric reporting method, model and hyperparameters (includig data augmentation scheduling) as in the Stanford 2D FSE experiments in Section \ref{appendix:stanford2d}. We scale all original measurements by $10^{-6}$.


\begin{figure}[htb]
	\captionsetup[subfigure]{labelformat=empty}
	\centering 
	\begin{subfigure}{0.24\textwidth}
		\includegraphics[width=\linewidth]{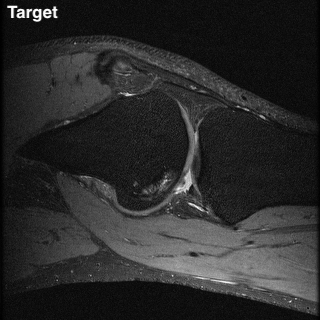}
		\label{fig:s3d1}
		\vspace{-0.2cm}
	\end{subfigure}\hfil 
	\begin{subfigure}{0.24\textwidth}
		\includegraphics[width=\linewidth]{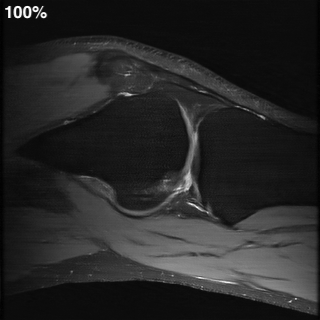}
		\label{fig:s3d2}
		\vspace{-0.2cm}
	\end{subfigure}\hfil 
	\begin{subfigure}{0.24\textwidth}
		\includegraphics[width=\linewidth]{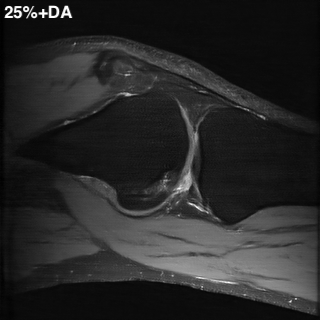}
		\label{fig:s3d9}
		\vspace{-0.2cm}
	\end{subfigure}\hfil
	\begin{subfigure}{0.24\textwidth}
		\includegraphics[width=\linewidth]{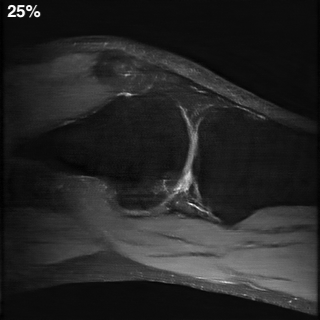}
		\label{fig:s3d4}
		\vspace{-0.2cm}
	\end{subfigure}\hfil 
	\caption{Visual comparison of reconstructions on the Stanford Fullysampled 3D FSE Knees dataset under various amount of training data, with and without data augmentation.}
	\label{fig:stanford3d_vis}
\end{figure}


\section{Robustness experiments\label{appendix:robustness}}

%

\textbf{Validation on unseen MRI scanners.} We explore how data augmentation impacts generalization to new MRI scanner models not available in training time. Different MRI scanners may use different field strenghts for acquisition, and higher field strength typically correlates with higher SNR. Volumes in the fastMRI knee dataset have been acquired by the following $4$ different scanners (followed by field strength): \textit{MAGNETOM Aera (1.5T), MAGNETOM Skyra (3T),  Biograph mMR (3T)} and \textit{MAGNETOM Prisma Fit (3T)}. The number of slices acquired by the different scanners are shown in Table \ref{tab:scanner_slices}. We perform the following experiments:
\begin{itemize}
	\item $3T \rightarrow 3T$: We train and validate on volumes acquired by scanners with $3T$ field strength. The training set only contains scans from \textit{MAGNETOM Skyra} and \textit{MAGNETOM Prisma Fit}, and we validate on \textit{Biograph mMR} scans. 
	\item $3T \rightarrow 1.5T$: We train on all volumes acquired by $3T$ scanners and validate on the $1.5T$ scanner (\textit{MAGNETOM Aera}).
	\item $1.5T \rightarrow 3T$: We train on all volumes acquired by the $1.5T$ scanner and validate on all other $3T$ scanners.
\end{itemize}
We combine volumes corresponding to the same scanner model from the official train and validation sets to form the validation set, however for training we only use volumes of the given model from the training set. Table \ref{tab:scanner_transfer} summarizes our results. 

\begin{table}[t]
	\begin{minipage}{0.47\textwidth}
		\centering
		\resizebox{7.7cm}{!}{
			\begin{tabular}{|c|c|c|}
				\hline
				\textbf{Model} & \textbf{Slices in train}& \textbf{Slices in val} \\ \hline
				Aera (1.5T) &  $13856$  &$3200$ \\ 
				Skyra (3T) &  $ 15370$ & $2967$ \\
				Biograph (3T) & $3961$ & $748$ \\
				Prisma Fit (3T)&  $1555$ & $220$ \\ \hline
			\end{tabular}
		}
		\caption{\label{tab:scanner_slices} Number of available slices for each scanner type in the train and validation splits of the fastMRI dataset. 		\vspace{0.2cm}} 
		
		\resizebox{7.7cm}{!}{
			\begin{tabular}{|l|c|c|}
				\hline
				\textbf{Transform} & \textbf{Range of values} & $w_i$\\ \hline
				H-flip&  flipped/not flipped & 0.5 \\ 
				V-flip& flipped/not flipped & 0.5 \\ 
				Rot. by $k \cdot 90^\circ$&  $k \in \{0, 1, 2, 3\}$ & 0.5 \\ 
				Rotation & $ [-180^\circ, 180^\circ]$ & 0.5 \\ 
				Translation &  \small{x: $ [-8\%, 8\%]$, y: $[-12.5\%, 12.5\%]$ }& 1.0\\ 
				Iso. scaling &  $[0.75, 1.25]$ & 0.5\\ 
				Aniso. scaling &  $[0.75, 1.25]$ along each axes & 0.5\\ 
				Shearing &  $ [-12.5^\circ, 12.5^\circ]$ & 1.0\\ \hline
			\end{tabular}
		}
		\caption{\label{tab:da_params}  Data augmentation configuration for all fastMRI experiments.}
	\end{minipage}
	\hfill
	\begin{minipage}{0.47\textwidth}
		\centering
		\begin{tabular}{|l|c|c|}
			\hline
			\textbf{Augmentations} & \textbf{SSIM} \\ \hline
			none&  $0.8396$  \\ 
			pixel preserving only &  $ 0.8585$ \\
			interpolating only &0.8731 \\
			all augmentations &  \textbf{ 0.8758} \\ \hline
		\end{tabular}
		\caption{\label{tab:ablation_transform} Comparison of peak validation SSIM applying various sets of augmentations on $1\%$ of fastMRI training data, multi-coil acquisition.\vspace{0.5cm}} 
		\centering
		\begin{tabular}{|l|c|c|}
			\hline
			\textbf{Augmentation scheduling} & \textbf{SSIM} \\ \hline
			none&  $0.8396$  \\ 
			exponential, $0.3$ &  $ 0.8565$ \\
			constant, $0.3$  &$0.8588$ \\
			exponential, $0.6$ &  \textbf{ 0.8758} \\ 
			constant, $0.6$ &  $0.8611$ \\ 
			exponential, $0.8$ &  $0.8600$ \\ \hline
		\end{tabular}
		\caption{\label{tab:ablation_schedule} Comparison of peak validation SSIM using different augmentation probability schedules on $1\%$ of fastMRI training data, multi-coil acquisition.}
	\end{minipage}
\end{table}

\section{Ablation studies \label{appendix:ablation}}
\textbf{Transformations.} We performed ablation studies on $1\%$ of the fastMRI knee training dataset in order to better understand which augmentations are useful. We use the multi-coil experiment with all augmentations as baseline and tune augmentation probability for other experiments such that the probability that a certain slice is augmented by at least one augmentation is the same across all experiments. We depict results on the validation dataset in Table \ref{tab:ablation_transform}. Both pixel preserving and general (interpolating) affine transformations are useful and can significantly increase reconstruction quality. Furthermore, we observe that their effect is complementary: they are helpful separately, but we achieve peak reconstruction SSIM when all applied together. Finally, the utility of pixel preserving augmentations seems to be lower than that of general affine augmentations, however they come with a negligible additional computational cost.

\textbf{Augmentation scheduling.} Furthermore, we investigate the effect of varying the augmentation probability scheduling function. The results on the validation dataset are depicted in Table \ref{tab:ablation_schedule}, where \textit{exponential, $\hat{p}$} denotes the exponential scheduling function in
\begin{equation*}
	p(t) = \frac{p_{max}}{1-e^{-c}}(1-e^{-t c/T}),
\end{equation*}
with $p_{max}=\hat{p} $ and \textit{constant, $\hat{p}$} means we use a fixed augmentation probability $\hat{p}$ throughout training. We observe that scheduling starting from low augmentation probability and gradually increasing is better than a constant probability, as initially the network does not benefit much from data augmentation as it can still learn from the original samples. Furthermore, too low or too high augmentation probability both degrade performance. If the augmentation probability is too low, the network may overfit to training data as more regularization is needed. On the other hand, too much data augmentation hinders reconstruction performance as the network rarely sees images close to the original training distribution.

\onecolumn

\begin{figure}[htb]
	\captionsetup[subfigure]{labelformat=empty}
	\centering 
	\begin{subfigure}{0.242\textwidth}
		\includegraphics[width=\linewidth]{plots/stanford2d/reconstructions/target_1.png}
		\label{fig:s2d1}
		\vspace{-0.30cm}
	\end{subfigure}\hfil 
	\begin{subfigure}{0.242\textwidth}
		\includegraphics[width=\linewidth]{plots/stanford2d/reconstructions/recon_train100_noDA_seed3_1.png}
		\label{fig:s2d2}
		\vspace{-0.30cm}
	\end{subfigure}\hfil 
	\begin{subfigure}{0.242\textwidth}
		\includegraphics[width=\linewidth]{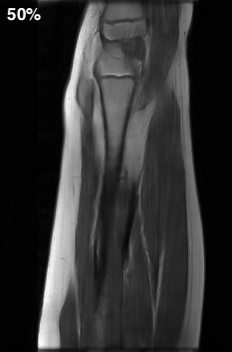}
		\label{fig:s2d3}
		\vspace{-0.30cm}
	\end{subfigure}\hfil 
	\begin{subfigure}{0.242\textwidth}
		\includegraphics[width=\linewidth]{plots/stanford2d/reconstructions/recon_train25_noDA_seed3_1.png}
		\label{fig:s2d4}
		\vspace{-0.30cm}
	\end{subfigure}\hfil 
	\begin{subfigure}{0.242\textwidth}
		\includegraphics[width=\linewidth]{plots/stanford2d/reconstructions/target_1.png}
		\label{fig:s2d6}
	\end{subfigure}\hfil 
	\begin{subfigure}{0.242\textwidth}
		\includegraphics[width=\linewidth]{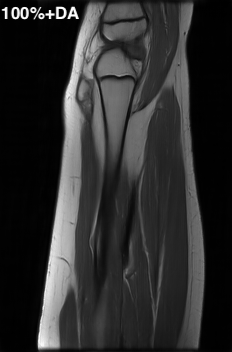}
		\label{fig:s2d7}
	\end{subfigure}\hfil 
	\begin{subfigure}{0.242\textwidth}
		\includegraphics[width=\linewidth]{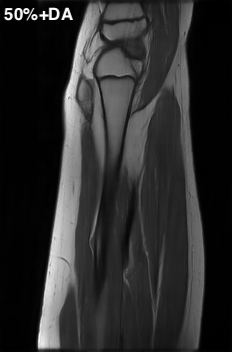}
		\label{fig:s2d8}
	\end{subfigure}\hfil 
	\begin{subfigure}{0.242\textwidth}
		\includegraphics[width=\linewidth]{plots/stanford2d/reconstructions/recon_train25_DA_seed3_1.png}
		\label{fig:s2d9}
	\end{subfigure}
	\begin{subfigure}{0.242\textwidth}
		\includegraphics[width=\linewidth]{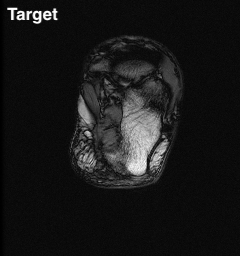}
		\vspace{-0.30cm}
	\end{subfigure}\hfil 
	\begin{subfigure}{0.242\textwidth}
		\includegraphics[width=\linewidth]{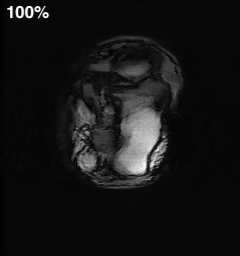}
		\vspace{-0.30cm}
	\end{subfigure}\hfil 
	\begin{subfigure}{0.242\textwidth}
		\includegraphics[width=\linewidth]{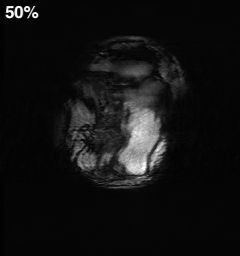}
		\vspace{-0.30cm}
	\end{subfigure}\hfil 
	\begin{subfigure}{0.242\textwidth}
		\includegraphics[width=\linewidth]{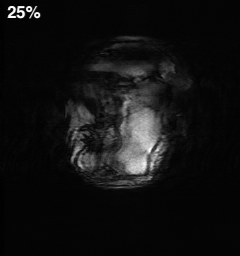}
		\vspace{-0.30cm}
	\end{subfigure}\hfil 
	\begin{subfigure}{0.242\textwidth}
		\includegraphics[width=\linewidth]{plots/stanford2d/reconstructions/target_2.png}
	\end{subfigure}\hfil 
	\begin{subfigure}{0.242\textwidth}
		\includegraphics[width=\linewidth]{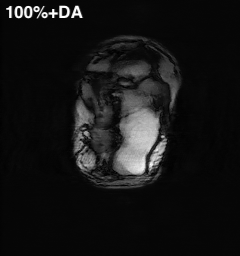}
	\end{subfigure}\hfil 
	\begin{subfigure}{0.242\textwidth}
		\includegraphics[width=\linewidth]{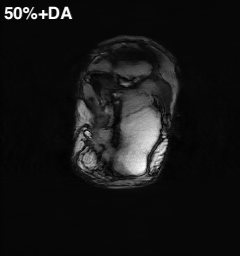}
	\end{subfigure}\hfil 
	\begin{subfigure}{0.242\textwidth}
		\includegraphics[width=\linewidth]{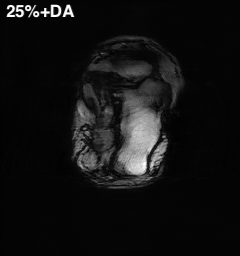}
	\end{subfigure}
	\caption{Visual comparison of reconstructions on the Stanford 2D FSE dataset under various amount of training data, with and without data augmentation.}
	\label{fig:stanford2d_vis_sup}
\end{figure}
\end{document}